\documentclass[sigconf]{acmart}

\usepackage{amsmath,amsfonts,bm}

\def\eqref#1{equation~\ref{#1}}

\def\1{\bm{1}}

\DeclareMathAlphabet{\mathsfit}{\encodingdefault}{\sfdefault}{m}{sl}
\SetMathAlphabet{\mathsfit}{bold}{\encodingdefault}{\sfdefault}{bx}{n}

\usepackage{amsmath}
\usepackage{pifont}
\usepackage{makecell}
\usepackage[most]{tcolorbox}
\usepackage{shadowtext}
\usepackage{censor}
\usepackage{xspace}
\usepackage{url}
\usepackage{booktabs}
\usepackage{nicefrac}
\usepackage{microtype}
\usepackage{setspace}
\usepackage[boxed, ruled, vlined, linesnumbered]{algorithm2e}
\usepackage{algorithmic}
\usepackage{wrapfig}
\usepackage{graphicx}
\usepackage{multirow}
\usepackage{colortbl}
\usepackage{soul}
\usepackage[colorinlistoftodos]{todonotes}
\usepackage{multicol}
\usepackage{xcolor}
\usepackage{pdfpages}
\usepackage{marvosym}
\usepackage{cleveref}

\usepackage{amssymb}
\usepackage{float}
\usepackage{subcaption}
\usepackage{fontawesome}
\usepackage{enumitem}

\definecolor{cellcolor}{RGB}{0,0,0}
\definecolor{antiquewhite}{rgb}{0.98, 0.92, 0.84}
\definecolor{anti-flashwhite}{rgb}{0.95, 0.95, 0.96}
\definecolor{aliceblue}{rgb}{0.94, 0.97, 1.0}
\definecolor{almond}{rgb}{0.94, 0.87, 0.8}
\definecolor{cosmiclatte}{rgb}{1.0, 0.97, 0.91}
\definecolor{darkbyzantium}{rgb}{0.36, 0.22, 0.33}
\definecolor{darkseagreen}{rgb}{0.56, 0.74, 0.56}
\definecolor{darkspringgreen}{rgb}{0.09, 0.45, 0.27}
\definecolor{asparagus}{rgb}{0.53, 0.66, 0.42}
\definecolor{antiquefuchsia}{rgb}{0.57, 0.36, 0.51}
\definecolor{ao(english)}{rgb}{0.0, 0.5, 0.0}
\definecolor{deepcerise}{rgb}{0.85, 0.2, 0.53}
\definecolor{denim}{rgb}{0.08, 0.38, 0.74}
\definecolor{crimson}{rgb}{0.86, 0.08, 0.24}
\definecolor{buff}{rgb}{0.94, 0.86, 0.51}
\definecolor{amber(sae/ece)}{rgb}{1.0, 0.49, 0.0}
\definecolor{airforceblue}{rgb}{0.36, 0.54, 0.66}
\definecolor{amethyst}{rgb}{0.6, 0.4, 0.8}
\definecolor{azure(colorwheel)}{rgb}{0.0, 0.5, 1.0}
\definecolor{azure(web)(azuremist)}{rgb}{0.94, 1.0, 1.0}
\definecolor{beige}{rgb}{0.96, 0.96, 0.86}
\definecolor{cornsilk}{rgb}{1.0, 0.97, 0.86}
\definecolor{darkcerulean}{rgb}{0.03, 0.27, 0.49}
\definecolor{babyblue}{rgb}{0.54, 0.81, 0.94}
\definecolor{ballblue}{rgb}{0.13, 0.67, 0.8}
\definecolor{dukeblue}{rgb}{0.0, 0.0, 0.61}
\definecolor{champagne}{rgb}{0.97, 0.91, 0.81}
\definecolor{antiquefuchsia}{rgb}{0.57, 0.36, 0.51}
\definecolor{darkgreen}{rgb}{0, 0.5, 0}

\definecolor{cellshade}{HTML}{F0EDFF}

\usepackage{tikz}
\usetikzlibrary{positioning}
\usetikzlibrary{calc,positioning,fit}

\newcommand{\mycode}{\url{https://anonymous.4open.science/r/CTI-Gap-3CBA}\xspace}

\newcommand{\cmark}{\textcolor{darkgreen}{\ding{51}}}
\newcommand{\xmark}{\textcolor{red}{\ding{55}}}

\newtcolorbox{bluecolorbox}{
  colback=blue!3,
  colframe=blue!75!black,
  coltitle=black,
  fonttitle=\bfseries,
  boxrule=1.5pt,
  arc=3pt,
  left=2pt, right=2pt, top=1pt, bottom=1pt
}

\newtcolorbox{darkbluecolorbox}{
  colback=darkcerulean!5,
  colframe=darkcerulean!100,
  coltitle=black,
  fonttitle=\bfseries,
  boxrule=1.5pt,
  arc=3pt,
  left=2pt, right=2pt, top=1pt, bottom=1pt
}

\newtcolorbox{yellowcolorbox}{
  colback=yellow!7,
  colframe=orange!60,
  coltitle=black,
  fonttitle=\bfseries,
  boxrule=1.5pt,
  arc=3pt,
  left=2pt, right=2pt, top=1pt, bottom=1pt
}

\newcommand{\lightmidrule}{\arrayrulecolor{gray!50}\midrule\arrayrulecolor{black}}

\newcommand{\placeholder}[1]{\texttt{<\!#1\!>}}
\newtcolorbox{promptbox}{
  breakable,
  colback=blue!1,
  colframe=blue!40!black,
  boxrule=0.8pt,
  arc=2pt,
  left=6pt, right=6pt, top=6pt, bottom=6pt,
  title style={font=\bfseries},
  fonttitle=\bfseries
}

\newtcolorbox{greencolorbox}{
  colback=green!3,  
  colframe=green!70!black, 
  coltitle=black,        
  fonttitle=\bfseries,   
  boxrule=1.5pt,         
  arc=3pt,               
  left=2pt, right=2pt, top=1pt, bottom=1pt 
}

\newtcolorbox{legendbox}[1]{
    enhanced,
    colback=white,      
    colframe=black,     
    coltitle=black,     
    fonttitle=\itshape\large, 
    title={#1},
    sharp corners,      
    boxrule=0.7pt,      
    attach boxed title to top left={xshift=10pt, yshift*=-\tcboxedtitleheight/2},
    boxed title style={
        colback=white,  
        frame hidden,   
        size=small,
        bottom=0pt, top=0pt 
    },
    top=-2pt,                  
    bottom=2pt,             
    left=3pt,              
    right=3pt,                
}

\newtcolorbox{casebox}[2][]{
    enhanced,
    colframe=dukeblue,   
    colback=babyblue!5,        
    coltitle=black,           
    fonttitle=\bfseries\small,
    title={#2},               
    attach boxed title to top left={xshift=5mm, yshift*=-3mm}, 
    boxed title style={
        colback=white,    
        sharp corners,        
        frame hidden,         
    },
    sharp corners,            
    boxrule=0.5mm,            
    top=-2pt,                  
    bottom=2pt,             
    left=3pt,              
    right=3pt,                
    width=\linewidth,         
    #1                        
}

\AtBeginDocument{%
  }

\setcopyright{none}
\renewcommand\footnotetextcopyrightpermission[1]{}
\begin{document}

\title{Uncovering Vulnerabilities of LLM-Assisted Cyber Threat Intelligence}

\author{
Yuqiao Meng$^{1}$, Luoxi Tang$^{1}$, Feiyang Yu$^{2}$, Jinyuan Jia$^{3}$, Guanhua Yan$^{1}$, Ping Yang$^{1}$, Zhaohan Xi$^{1}$\\
$^{1}$Binghamton University \quad
$^{2}$Duke University \quad
$^{3}$Pennsylvania State University\\
\texttt{\{ymeng15, ltang24, zxi1\}@binghamton.edu}
}


\begin{abstract}

Large language models (LLMs) are increasingly used to help security analysts manage the surge of cyber threats, automating tasks from vulnerability assessment to incident response. Yet in operational CTI workflows, reliability gaps remain substantial. Existing explanations often point to generic model issues (e.g., hallucination), but we argue the dominant bottleneck is the threat landscape itself: CTI is heterogeneous, volatile, and fragmented. Under these conditions, evidence is intertwined, crowdsourced, and temporally unstable, which are properties that standard LLM-based studies rarely capture.

In this paper, we present a comprehensive empirical study of LLM vulnerabilities in CTI reasoning. We introduce a human-in-the-loop categorization framework that robustly labels failure modes across the CTI lifecycle, avoiding the brittleness of automated ``LLM-as-a-judge'' pipelines. We identify three domain-specific cognitive failures: spurious correlations from superficial metadata, contradictory knowledge from conflicting sources, and constrained generalization to emerging threats. We validate these mechanisms via causal interventions and show that targeted defenses reduce failure rates significantly. Together, these results offer a concrete roadmap for building resilient, domain-aware CTI agents.
\end{abstract}

\ccsdesc[500]{Security and privacy}
\ccsdesc[500]{Security and privacy~Vulnerability management}
\ccsdesc[300]{Computing methodologies~Natural language processing}
\ccsdesc[300]{Computing methodologies~Machine learning}

\keywords{large language models, cyber threat intelligence, vulnerability analysis, LLM security, CTI}


\maketitle
\pagestyle{plain}

\section{Introduction}
\label{sec:intro}

We are living in an era of rapid digital transformation, where technological advancements are tightly associated with the growing prevalence of cyber threats. In recent years, the cyber threat landscape has shifted dramatically, with reported Common Vulnerabilities and Exposures (CVEs) increasing by an average of 25\% annually \citep{cve}. In 2025 alone, more than 48,000 vulnerabilities were reported. This surge can be attributed to the rising complexity of IT systems \citep{stalnaker2024boms}, the widespread adoption of open-source software \citep{dam2023towards}, and the accelerating pace of modern development cycles \citep{trinkenreich2025get}. Together, these dynamics expand the attack surface while making sole reliance on human analysts for vulnerability assessment and remediation increasingly infeasible.

Large language models (LLMs) have recently demonstrated strong performance in a broad range of cyber threat intelligence (CTI) tasks. By adapting models through instruction fine-tuning or prompt-based automation, researchers have applied LLMs to support threat analysis and decision-making \citep{zhang2023hackmentor}, code vulnerability detection \citep{du2024generalization}, and defense against network intrusions \citep{lavi2024fine}. Despite these advances, substantial performance gaps are reported in their  evaluations \citep{deng2024pentestgpt,clairoux2024use,ji2024sevenllm,alam2024ctibench,liu2025benchmarking}, suggesting that such limitations cannot be fully addressed through typical model adaptation or prompt-based automation. This raises a fundamental research question: \textbf{What intrinsic vulnerabilities constrain the effectiveness of LLMs in supporting CTI tasks?}

\begin{figure}[t]
    \centering
    \includegraphics[width=\linewidth]{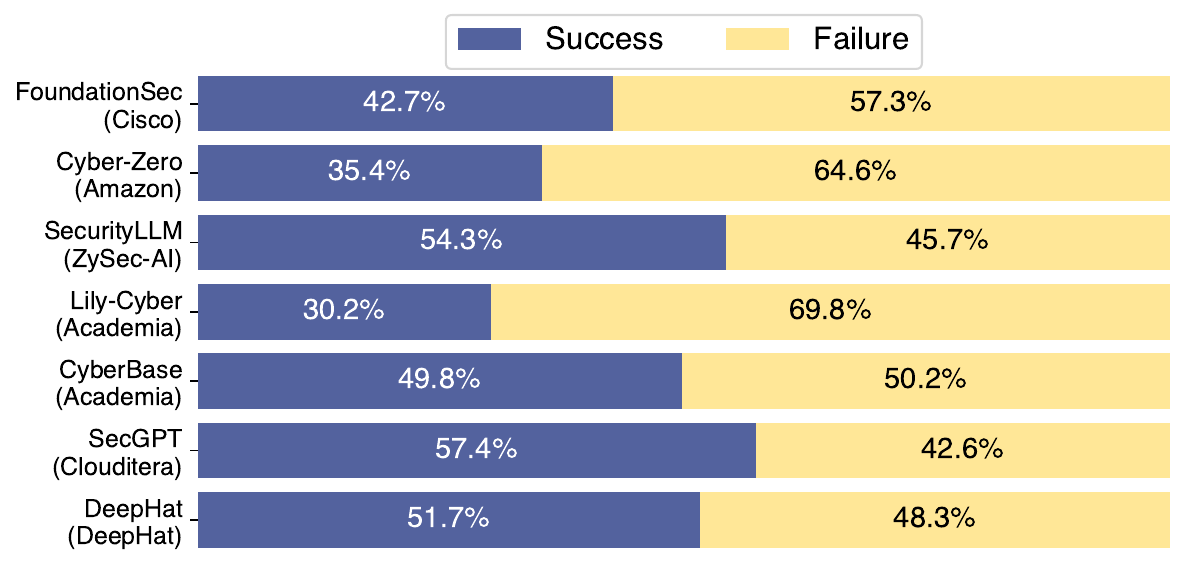}
    \caption{Failure ratios of LLM-based cybersecurity agents on large-scale CTI tasks (detailed evaluations in \S\ref{ssec:eval}).}
    \label{fig:example}
    
\end{figure}

\begin{figure*}[t]
    \centering
    \includegraphics[width=\linewidth]{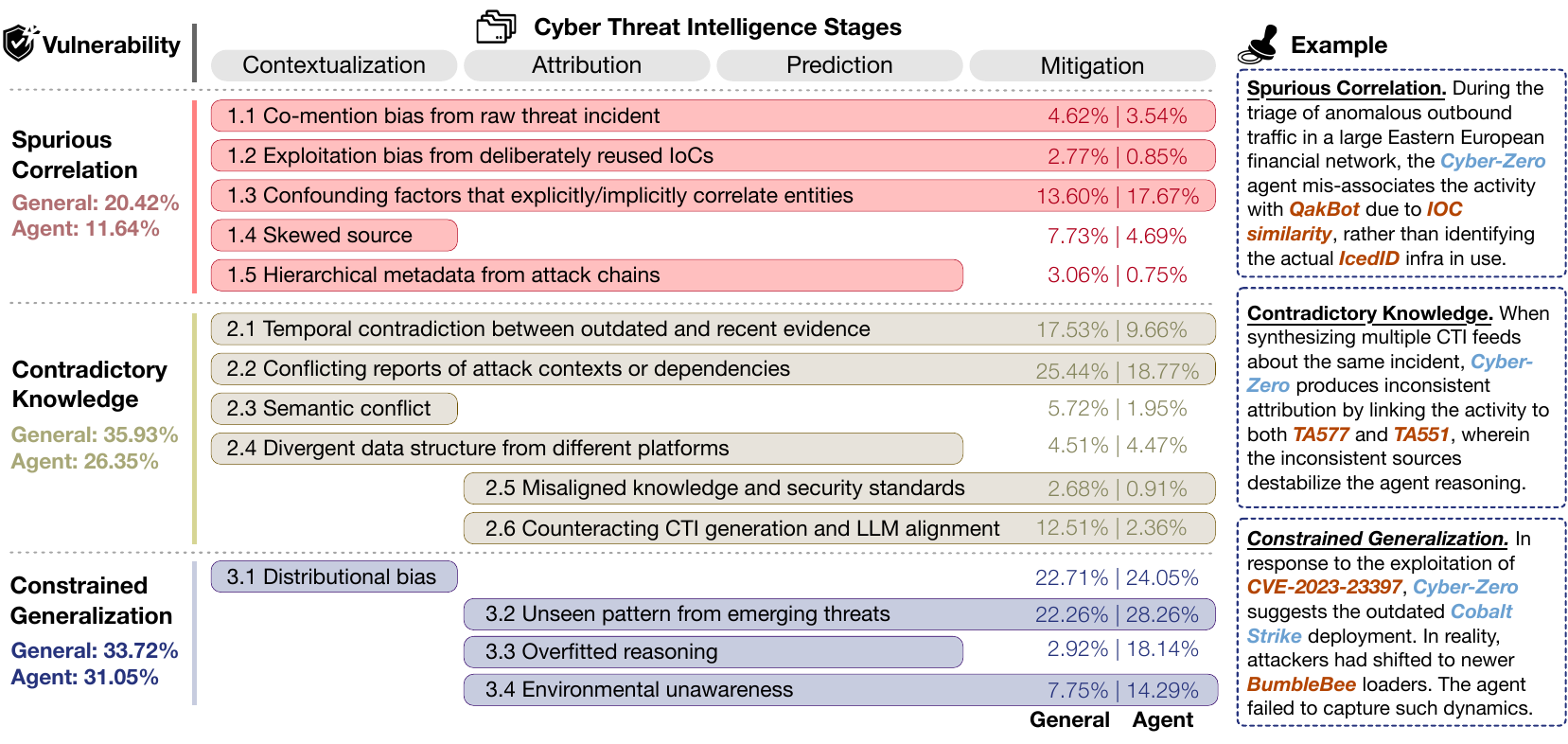}
    \caption{Summarization and examples of vulnerability types of LLMs in various CTI stages. Ratios are calculated over the entire dataset (\S\ref{ssec:eval}). \textbf{Several vulnerabilities may present simultaneously in a single threat instance.}}
    \label{fig:summary}
\end{figure*}

Although some vulnerabilities (e.g., hallucination \citep{ji2023towards,martino2023knowledge,friel2023chainpoll},  jailbreaking \citep{niu2024jailbreaking,coalson2024prisonbreak}) arise from general deficiencies in LLM architecture and training methodologies that affect broad domains \citep{aguilera2025llm}, studying these limitations in a domain-agnostic manner fails to propose effective, cyber-centric solutions. Notably, developing LLM-based CTI agents requires navigating the threat landscape defined by the heterogeneity and volatility of its knowledge sources \cite{rigaki2024hackphyr,de2025cfa}. Unlike static corpora, CTI is a fragmented ecosystem comprising structured databases (e.g., NVD \citep{nvd}), semi-structured vendor advisories \citep{berndtsson2017combining}, and unstructured discussions on social media or dark web forums \citep{wang2019automatic,rahman2020literature}. Furthermore, cyber threats are ever-evolving; the severity or remediation status of a vulnerability is subject to rapid change based on real-time exploit availability \citep{pureti2022zero}. Consequently, standard LLM evaluations often fail to capture the complexity of the disparate, time-sensitive, and frequently contradictory threat landscape.

\noindent
\textbf{This work.} Given that developing CTI agents requires reasoning under intertwined, crowdsourced, and imbalanced evidence, we aim to study: (1) which vulnerabilities are uniquely derived from the threat landscape itself; (2) how these factors introduce specific weaknesses in LLM-assisted CTI; and (3) what defensive solutions can effectively mitigate them. Specifically, we make the following contributions:

\begin{figure*}[t] 
    \begin{legendbox}{Example: CTI stages}
        A healthcare network detects suspicious outbound traffic linked to newly registered domains associated with \textit{QakBot}. \textbf{The security team first enriches event metadata} by mapping proxy logs to known \textit{C2 infrastructure} and retrieving prior reports for context. \textbf{Next, they attribute} \textit{QakBot}’s reuse of infrastructure to the threat group \textit{TA577}. Based on historical correlations, \textbf{they then predict} a likely transition to \textit{Cobalt Strike} and eventual ransomware deployment. Finally, \textbf{the team implements mitigation strategies}, including generating \textit{Sigma} detection rules and prioritizing patches for vulnerable \textit{Exchange servers}.
    \end{legendbox}
\end{figure*}

\underline{\textbf{I. A comprehensive empirical evaluation.}} To characterize LLM failure modes in CTI, we conduct a large-scale, systematic study spanning the complete CTI lifecycle, from threat analysis to incident response. Our evaluation leverages a diverse suite of datasets, integrating established benchmarks (CTIBench \citep{alam2024ctibench}, SevenLLM-Bench \citep{ji2024sevenllm}, SWE-Bench \citep{jimenez2023swe}, and CyberTeam \citep{liu2025benchmarking}) with real-world threat reports sourced from authoritative knowledge bases or threat feeds (e.g., \citep{cve,nvd,CISA_KEV_Catalog}) to ensure  comprehensiveness. Figure~\ref{fig:example} shows summarized failure ratios among cyber LLM agents.

\underline{\textbf{II. A failure categorization methodology.}} Based on our evaluation results, we develop a rigorous methodology to investigate and categorize failure instances. A critical challenge in analyzing extensive threat data is reliably scaling the classification of diverse failure modes. Standard LLM-as-a-judge approaches prove unreliable for automatic failure classification, as models tend to rationalize their own outputs or fail to critically evaluate contradictions \citep{yamauchi2025empirical,guerdan2025validating}. To address this, we propose an autoregressive, human-in-the-loop framework that categorizes failure instances with high efficiency and reliability (\S\ref{sec:method}). This methodology enables us to extract data-grounded insights from large-scale evaluations while significantly reducing the manual annotation workload.

\ul{\textbf{III. A taxonomy of CTI vulnerabilities and root cause analysis.}} Through large-scale experiments on industry-leading LLMs and LLM-powered cybersecurity agents, complemented by detailed case studies, we identify three dominant categories of vulnerabilities that limit LLM reasoning:
\textbf{(1) Spurious correlations}, wherein LLMs over-attribute based on superficial or co-occurring features (e.g., mistaking commodity tools as actor-specific evidence). 
\textbf{(2) Contradictory knowledge}, wherein inconsistencies in CTI sources confuse models, causing unstable or conflicting predictions.
\textbf{(3) Constrained generalization}, wherein LLMs struggle to extend beyond familiar distributions, failing on emerging (zero-day) attack surfaces.
    
Our analysis further shows that these vulnerabilities directly undermine the effectiveness of techniques used in CTI pipelines. As summarized in Figure~\ref{fig:summary}, each vulnerability spans multiple CTI stages and influences different aspects of LLM reasoning. For example, spurious correlations distort evidence retrieval, where LLMs amplify irrelevant co-occurrences in retrieved evidence, resulting in misguided contextualization and faulty attribution. 
Similarly, contradictory knowledge weakens entity recognition and event extraction, as models produce inconsistent mappings of the same threat actor or IOC across conflicting sources, thereby breaking downstream attack chain construction. 
Overall, the detailed studies reveal the root causes by combining vulnerabilities with the specific LLM workflows involved in different CTI stages, which provides an in-depth understanding of how these vulnerabilities propagate through the reasoning.

\ul{\textbf{IV. Validation and defensive strategies.}} We further validate the root causes of these vulnerabilities through a causal intervention study, showing that reasoning-time interventions can yield immediate performance gains. Building on these findings, we design and evaluate targeted mitigation strategies, including retrieval-based CTI knowledge filtering and inference-time reasoning constraints. Our results show that these lightweight, taxonomy-informed defenses consistently reduce failure rates, demonstrating that robust LLM-based CTI analysis requires protections explicitly tailored to the heterogeneity of threat landscape.

In summary, this work uncovers blind spots in LLM vulnerabilities for CTI. We provide foundational insights and initial defensive efforts that can guide principled adaptations in future CTI-focused LLM systems. To facilitate follow-up research, we release codes at \mycode.

\section{Background and Motivation}
\label{sec:task}

\begin{table*}[t]
\centering
\footnotesize
\def\arraystretch{0.95}
\setlength{\tabcolsep}{8pt}
\caption{Collected cybersecurity benchmarks for LLM-CTI.}
\resizebox{\textwidth}{!}{
\label{tab:bench}
\begin{tabular}{llccccl}
\toprule
\multirow{2.5}{*}{\bf Benchmark} & \multirow{2.5}{*}{\bf Description} & \multicolumn{4}{c}{\bf Task Coverage} & \multirow{2.5}{*}{\bf Unique Feature} \\
\cmidrule(lr){3-6}
& & \ding{182} Cont. & \ding{183} Attr. & \ding{184} Pred. & \ding{185} Miti. & \\ 
\midrule
CTIBench \citep{alam2024ctibench} & General CTI tasks & \cmark & \cmark & \cmark & \xmark & Multi-choice questions (MCQ)  \\
SevenLLM-Bench \citep{ji2024sevenllm} & Report understanding & \cmark & \cmark & \cmark & \cmark & Synthetic instances, MCQ, QA\\
SWE-Bench \citep{jimenez2023swe} & Software bug fixing & \xmark & \cmark & \xmark & \cmark & Program analysis \& patching \\
CyberTeam \citep{liu2025benchmarking} & Blue-team threat hunting & \cmark & \cmark & \cmark & \cmark & Open-ended decision-making \\
\bottomrule
\end{tabular}
}
\end{table*}
\begin{table*}[!t]
  \footnotesize
  \def\arraystretch{0.85}
  \setlength{\tabcolsep}{5pt}
  \caption{Evaluation of LLMs on CTI tasks across four CTI stages. Industry-leading general-purpose LLMs (left) are compared with cybersecurity-specialized models (right). Detailed CTI task descriptions (with examples) are provided in \ref{app:task-desc}. Model names and versions are introduced in  \ref{app:model-version}.}
  \centering
  \resizebox{\textwidth}{!}{%
  \begin{tabular}{ll|cccccccc|ccccccc}
    \toprule
    \multirow{2.5}{*}{\bf Detailed CTI Task} & \multirow{2.5}{*}{\bf Metric} & \multicolumn{8}{c}{\bf Industry-leading, general-purpose} & \multicolumn{7}{c}{\bf Cybersecurity-specialized} \\
    \cmidrule(lr){3-10}\cmidrule(lr){11-17}
     & & G5 & Go4 & CLD & GEM & LL70 & MIX & QWN & GRK & FSC & CB0 & ZYS & LLY & CBS & SPT & DHT \\
    \lightmidrule
    \multicolumn{17}{c}{\bf \ding{182} Contextualization} \\
    \lightmidrule
    Affected Systems           & F1   & .822 & .801 & .757 & .613 & .882 & .663 & .747 & .819 & .432 & .418 & .553 & .566 & .562 & .554 & .559 \\
    Attack Infrastructure      & F1   & .863 & .741 & .614 & .578 & .853 & .628 & .616 & .493 & .507 & .492 & .612 & .625 & .618 & .611 & .616 \\
    Vulnerability Linking      & Acc  & .652 & .633 & .602 & .574 & .649 & .533 & .521 & .497 & .512 & .496 & .621 & .633 & .629 & .622 & .628 \\
    Malware Family Mapping     & F1   & .681 & .659 & .635 & .602 & .584 & .567 & .551 & .529 & .541 & .526 & .639 & .652 & .646 & .641 & .648 \\
    IOC Normalization          & F1   & .721 & .707 & .682 & .661 & .642 & .623 & .609 & .593 & .602 & .589 & .678 & .689 & .684 & .678 & .683 \\
    Threat Report Alignment    & BLEU  & .429 & .218 & .206 & .396 & .486 & .279 & .271 & .363 & .159 & .352 & .441 & .348 & .244 & .139 & .342 \\
    Event Timeline Construction& BLEU & .563 & .549 & .532 & .519 & .504 & .492 & .478 & .468 & .472 & .459 & .594 & .602 & .599 & .593 & .597 \\
    Graph Population           & Acc  & .793 & .676 & .559 & .539 & .724 & .507 & .693 & .478 & .487 & .472 & .421 & .334 & .629 & .424 & .628 \\
    Source Reliability Scoring & AUC  & .912 & .894 & .773 & .861 & .745 & .631 & .717 & .753 & .712 & .699 & .738 & .547 & .642 & .537 & .641 \\
    \lightmidrule
    \multicolumn{17}{c}{\bf \ding{183} Attribution} \\
    \lightmidrule
    Threat Actor Linking       & Acc  & .892 & .871 & .652 & .822 & .598 & .773 & .753 & .528 & .643 & .526 & .704 & .413 & .707 & .771 & .608 \\
    TTP Extraction             & F1   & .751 & .738 & .724 & .703 & .478 & .669 & .654 & .642 & .654 & .639 & .724 & .537 & .731 & .726 & .732 \\
    Campaign Attribution       & Acc  & .712 & .691 & .671 & .649 & .631 & .607 & .586 & .567 & .578 & .563 & .694 & .703 & .699 & .694 & .701 \\
    Infrastructure Reuse       & F1   & .677 & .656 & .636 & .609 & .591 & .574 & .556 & .534 & .548 & .531 & .688 & .528 & .692 & .754 & .603 \\
    Language/Style Profiling   & Acc  & .598 & .581 & .561 & .539 & .521 & .503 & .488 & .475 & .489 & .476 & .632 & .643 & .638 & .633 & .639 \\
    False Flag Detection       & F1  & .679 & .526 & .501 & .486 & .672 & .459 & .547 & .436 & .444 & .431 & .574 & .286 & .462 & .576 & .582 \\
    Evidence Weighting         & BLEU  & .362 & .247 & .131 & .226 & .402 & .288 & .178 & .207 & .073 & .059 & .097 & .007 & .102 & .197 & .083 \\
    Relation Graph Building    & F1   & .642 & .628 & .611 & .595 & .579 & .562 & .547 & .533 & .544 & .528 & .675 & .683 & .678 & .673 & .679 \\
    \lightmidrule
    \multicolumn{17}{c}{\bf \ding{184} Prediction} \\
    \lightmidrule
    Exploit Likelihood         & AUC  & .821 & .806 & .792 & .771 & .856 & .742 & .629 & .714 & .519 & .703 & .742 & .559 & .764 & .653 & .759 \\
    Impact Forecast            & BLEU  & .498 & .383 & .271 & .354 & .441 & .226 & .213 & .202 & .108 & .094 & .207 & .119 & .046 & .115 & .221 \\
    Target Sector Prediction   & Acc  & .841 & .759 & .743 & .819 & .602 & .756 & .712 & .553 & .564 & .679 & .502 & .311 & .623 & .415 & .619 \\
    Campaign Escalation        & AUC  & .683 & .627 & .611 & .598 & .582 & .568 & .554 & .541 & .548 & .532 & .607 & .518 & .652 & .607 & .613 \\
    \lightmidrule
    \multicolumn{17}{c}{\bf \ding{185} Mitigation} \\
    \lightmidrule
    Patch Recommendation       & F1  & .702 & .679 & .659 & .636 & .718 & .601 & .583 & .671 & .582 & .567 & .632 & .442 & .629 & .641 & .446 \\
    Rule Generation (YARA)     & BLEU   & .382 & .216 & .339 & .482 & .267 & .213 & .337 & .184 & .231 & .307 & .281 & .089 & .094 & .287 & .202 \\
    Response Summarization     & BLEU  & .514 & .399 & .286 & .567 & .252 & .236 & .422 & .311 & .118 & .304 & .315 & .227 & .433 & .126 & .208 \\
    Mitigation–TTP Mapping     & Acc  & .672 & .652 & .633 & .611 & .596 & .578 & .561 & .548 & .559 & .544 & .613 & .626 & .631 & .624 & .629 \\
    Defensive Playbook Gen     & BLEU & .586 & .572 & .557 & .537 & .522 & .508 & .495 & .482 & .491 & .476 & .561 & .572 & .576 & .571 & .575 \\
    Countermeasure Ranking     & NDCG & .591 & .574 & .561 & .547 & .532 & .519 & .504 & .494 & .503 & .489 & .623 & .533 & .629 & .424 & .429 \\
    Incident Ticket Generation & Acc & .831 & .716 & .601 & .682 & .868 & .753 & .639 & .628 & .536 & .421 & .653 & .464 & .668 & .602 & .564 \\
    \bottomrule
  \end{tabular}
  }
  \label{tab:eval}
\end{table*}

This section first outlines the scope of CTI tasks (\S\ref{ssec:cti_stages}), then introduces the datasets used to assess LLM performance along with extensive evaluation results (\S\ref{ssec:eval}). These results motivate a deeper investigation into specific vulnerabilities and their root causes, which we address through our methodological design to categorize failure instances (\S\ref{sec:method})  and research findings (\S\ref{sec:expt}).


\subsection{Background: Stages of CTI Tasks}
\label{ssec:cti_stages}

Cyber threat intelligence (CTI) covers a broad range of cybersecurity activities that support the analysis of threat events and the recommendation of timely, informed incident response. 


As illustrated by the \textbf{example above}, CTI practices are typically organized into a pipeline consisting of four stages, each of which involves distinct reasoning tasks and technical solutions:

\ding{182} \textbf{Contextualization}: Security teams must enrich raw observations (e.g., suspicious logs, network alerts, isolated IOCs) with contextual information to make them actionable. This includes mapping events to known threat identifiers such as CVEs or MITRE ATT\&CK TTPs, linking indicators to malware families, and constructing coherent timelines of adversarial campaigns. \textbf{Involved techniques} in this stage include topic modeling to group related threat narratives, event extraction to identify structured incidents, knowledge base mapping to align content with known taxonomies, and information retrieval to ground outputs in relevant threat reports or databases.

\ding{183} \textbf{Attribution}: Once threat contexts are enriched, security teams investigate the likely adversaries behind the activity. Attribution connects threat events to specific actor profiles or campaigns by analyzing shared TTPs, infrastructure reuse (e.g., IP, domain overlap), and stylistic patterns such as language use or operational cadence. \textbf{Involved techniques} here include named entity recognition (NER) to extract actors, malware, and victim entities; relation extraction to identify links among entities and events; structured event graph construction to represent sequences of observed behavior; and threat actor classification using learned behavioral profiles from historical data.

\ding{184} \textbf{Prediction}: With an understanding of the adversary, security teams aim to forecast future threats, exploitation likelihood, and potential impact. This involves estimating the probability of exploitation for known vulnerabilities (e.g., EPSS scoring), anticipating which sectors or systems are likely to be targeted, and modeling campaign evolution. \textbf{Involved techniques} in this stage focus on historical event correlation to find temporal patterns, temporal modeling to capture threat progression, and forecasting using time series or graph neural networks to predict propagation or escalation risks.

\ding{185} \textbf{Mitigation}: Finally, CTI must support actionable decisions to reduce risk and guide incident response. This includes recommending specific patches, tuning detection signatures (e.g., YARA/Sigma rules), adjusting firewall or access control configurations, and drafting response playbooks. \textbf{Involved techniques} supporting this stage include mitigation mapping to associate observed TTPs or vulnerabilities with known defensive strategies, mitigation efficacy prediction to rank possible responses, and summarization to generate concise, structured remediation steps tailored to system environments. 

\begin{yellowcolorbox}
\underline{\textbf{Highlight \faLightbulbO.}} Involved techniques above not only define the operational CTI pipeline but also help explain the root causes of how vulnerabilities are triggered in cyberspace.
\end{yellowcolorbox}

Section \ref{ssec:root-cause} and Appendix~\ref{app:more-root-cause} analyze root causes of vulnerabilities triggered by these techniques across CTI stages, while Appendix~\ref{app:cti-stage} provides additional details on the techniques involved.

\subsection{Motivation: LLMs remain insufficient in variant CTI tasks}
\label{ssec:eval}

{\bf Evaluation Datasets.} To evaluate LLM performance on CTI tasks, we leverage  benchmarks \citep{alam2024ctibench,ji2024sevenllm,jimenez2023swe,liu2025benchmarking} as well as real-world threat databases and platforms. The benchmarks are summarized in Table \ref{tab:bench}, which provide a broad coverage across all CTI stages and capture both structured (MCQ, QA) and unstructured (decision-making, patch generation) task formats. To unify their use, we standardize each instance into a CTI-oriented scenario explicitly aligned with one of the four CTI stages. For example, multi-choice questions (MCQs) from CTIBench and SevenLLM-Bench are reformulated into concrete threat hunting scenarios, such as analyzing a suspicious log entry to determine the relevant TTP or linking an IOC to a known malware family. Our preprocessing mitigates structural biases that could otherwise inflate LLM performance.

We provide the details of real-world databases (or platforms) in Appendix~\ref{app:src-db}, along with data statistics in Table \ref{tab:eval-stats}. The prompt template used in evaluation is also included in Appendix \ref{app:eval-prompt}.

\begin{figure*}[t]
    \centering
    \includegraphics[width=\linewidth]{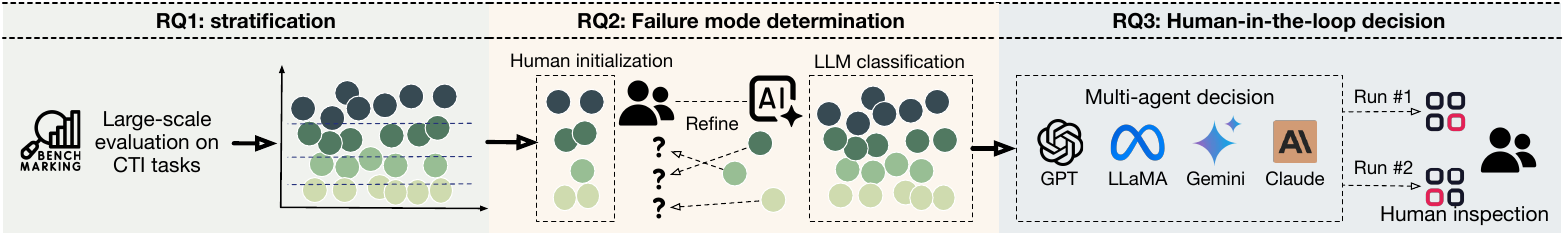}
    \caption{Overview of method to categorize failure instances (addressing RQ$_1$-RQ$_3$).}
    \label{fig:method}
\end{figure*}


{\bf Evaluated LLMs.} Our evaluation covers two complementary lines of models. First, we include industry-leading, general-purpose LLMs (e.g., GPT-5, Claude-Sonnet-4, Gemini-2.5), which represent state-of-the-art reasoning capabilities across domains. Second, we evaluate on open-source or API-accessible cybersecurity-specialized models (e.g., SecGPT \citep{SecGPT}, DeepHat \citep{DeepHat}), which are adapted to security operations through domain-specific training and curated CTI corpora. Investigating their performance gaps provides insights of vulnerabilities that cannot be fully addressed by either large-scale pretraining or cybersecurity-specialized adaptation. Appendix \ref{app:model-version} details all evaluated model and their versions.

\noindent\textbf{Model-specific Gap.} Across extensive evaluations on a broad range of CTI tasks (detailed introductions in \ref{app:task-desc}),  Table \ref{tab:eval} shows model-specific gaps: general-purpose LLMs dominate understanding-heavy and synthesis tasks (e.g., populating attack graphs), while the best cyber-specialized agents may stand out on semantic-driven or operational outputs (e.g., patch recommendation). In contextualization, general models post sizable wins, e.g., \emph{Affected Systems} F1 peaks at $\sim$0.82–0.88 versus $\sim$0.55–0.56 for most cyber agents; 
\emph{Source Reliability} AUC tops out at $\sim$0.91 versus a best cyber score of $\sim$0.74, which reflects constrained long-context retrieval and instruction-following that also present in text generation tasks such as \emph{Threat Report Alignment} and \emph{Event Timeline}. In attribution and forecasting, the trend largely holds: general LLMs lead \emph{Threat Actor Linking} (Acc up to $\sim$0.89) and \emph{Exploit Likelihood} (AUC up to $\sim$0.86). Notably, targeted domain tuning can flip certain edges: a top cyber agent outperforms on \emph{Infrastructure Reuse} (F1 $\sim$0.75 vs $\sim$0.68 best general), and cyber models consistently outperform general LLMs on operational ranking/classification such as \emph{Countermeasure Ranking} (NDCG $\sim$0.62–0.63 vs $\sim$0.57–0.59), suggesting influences from environmental (enterprise) contexts. Performance among cyber agents is also more uneven (e.g., low F1 on \emph{False Flag Detection} for some models), implying sensitivity to the quality and coverage of training data. 

{\bf Universal Gap.} We also observed universal gaps among all models: limited IOC normalization and CVE linking under obfuscation, inconsistent TTP extraction across retrieval-augmented reasoning, and weak temporal coherence in timelines/escalation forecasts. We also observe shallow reliance on real-world evidence in report alignment (low BLEU) and format errors in ticket/playbook generation. These deficiencies present with both larger online models (e.g., GPT-5) and cyber-specific agents, implying inherent limitations that may be triggered by the nature of the threat landscape.

Motivated by these gaps, we investigate vulnerabilities based on the concrete \textbf{failure modes} of LLMs in CTI tasks. We aim to build a systematic view about where models  break down and why these vulnerabilities persist despite large-scale pretraining or domain-specific adaptation.

\section{Methodology: Categorize Failure Instances}
\label{sec:method}

After conducting the large-scale evaluation described in \S\ref{sec:task}, we categorize the resulting failure cases to better understand the limitations of LLMs. This process is guided by three research questions: 

\noindent
\textbf{RQ$_1$:} How can we identify ``failure'' especially in tasks that lack hard-label annotations?

\noindent
\textbf{RQ$_2$:} How can we determine the finite scope of vulnerabilities (i.e., failure modes)?

\noindent
\textbf{RQ$_3$:} How can we efficiently categorize large-scale instances into these failure modes?

Overall, \textbf{we do not fully trust LLM-as-judge or AI-produced confidence} to detect failure cases, due to the lack of transparency and the risk of ``self-rationalized'' reasoning. Nevertheless, the large scale of the evaluation set requires us to categorize instances efficiently beyond purely manual efforts. To address this, we propose a \textbf{stratification} approach that partitions instances based on their ``failure'' depth, followed by an \textbf{autoregressive method} combined with \textbf{human-in-the-loop} efforts to resolve the above RQs. Figure \ref{fig:method} illustrates the overall workflow.

\subsection{Stratifying instances without hard-label annotations (RQ$_1$)} \label{ssec:strat}

\begin{algorithm}[t]
  \small 
  \SetAlgoLined
  \DontPrintSemicolon
  \caption{Stratified Failure Sampling (RQ$_1$)\label{algo:rq1}}
  
  \KwIn{Dataset $\mathcal{D}=\{x_i\}$, Models $\{y_i\}$, Refs $\{r_i\}$, Similarity measure $Sim(\cdot)$}
  \KwOut{Failure Set $\mathcal{F}$}

  \tcp{1. Score \& Stratify}
  Compute scores $\mathcal{S} = \{s_i \mid s_i = Sim(y_i, r_i), \forall i\}$\;
  Sort $\mathcal{S}$ and partition into quantile strata $\mathcal{Q} = \{Q_1, \dots, Q_M\}$ (bin size $\delta=0.05$)\;
  
  \tcp{2. Iterative Inspection}
  Initialize $\mathcal{F} \leftarrow \emptyset$, $Converged \leftarrow \text{False}$\;
  \ForEach{stratum $Q_k \in \mathcal{Q}$ \textbf{while} $\neg Converged$}{
      Sample anchors $A_k \subset Q_k$ for manual labeling\;
      Derive score ranges $R_{fail}, R_{pass}$ from labeled anchors $A_k$\;
      
      \tcp{Propagate labels to remaining $x_i$}
      \ForEach{$x_i \in Q_k \setminus A_k$}{
        \uIf{$s_i \in R_{fail}$}{
            $\mathcal{F} \leftarrow \mathcal{F} \cup \{x_i\}$ \;
        }
        \uElseIf{$s_i \in R_{pass}$}{
            \textbf{continue} \;
        }
        \Else{
            Manual check $x_i$; \textbf{if} failed $\mathcal{F} \leftarrow \mathcal{F} \cup \{x_i\}$ \tcp*{Boundary}
        }
      }
      
      \tcp{3. Termination Check}
      Update failure mode distribution $D_k$\;
      \If{$D_k \approx D_{k-1}$ \textbf{and} no new modes found}{
        $Converged \leftarrow \text{True}$\;
      }
  }
  \Return $\mathcal{F}$\;
\end{algorithm}

Most cyber threat intelligence tasks, such as mitigation rule generation, lack binary hard labels. However, authoritative reference materials (e.g., advisories from CISA \citep{CISA_KEV_Catalog} or CVE \citep{cve}) allow us to leverage reference-based similarity metrics (e.g., BLEU) to quantify the alignment between model outputs and ground truth.

To efficiently identify failure cases, we design a stratification approach (Algorithm~\ref{algo:rq1}). We first compute similarity scores for all instances and partition them into 5\% quantile-based strata. Within each stratum, we manually inspect a small set of anchor samples to establish score ranges for ``correct'' versus ``failed'' outputs. Remaining instances in the stratum are then automatically labeled based on these ranges; boundary cases falling in the overlapping region trigger additional manual review. This process proceeds iteratively through the strata and terminates once the distribution of failure modes stabilizes and no new modes emerge. Practically, this hybrid approach yields high-confidence failure sets while requiring manual inspection of less than 1.5\% of the total instances.

\subsection{Failure mode determination (RQ$_2$)}
\label{ssec:type}

We avoid using LLMs to directly determine failure modes (i.e., vulnerabilities), instead, we design an \textbf{iterative} process alternating between human annotation and LLM-assisted classification. Let $\mathcal{D}=\{x_i\}_{i=1}^N$ denote the set of failure instances (from stratification results in \S\ref{ssec:strat}).

\noindent
\underline{\textbf{Step 1 (Initialization).}} Human annotators randomly inspect a small subset $\mathcal{D}_0 \subset \mathcal{D}$ to derive an initial taxonomy of failure modes, $\mathcal{T}_0 = {t_1,\dots,t_k}$.

\noindent
\underline{\textbf{Step 2 (LLM classification).}} For each remaining instance $x_j \in \mathcal{D}\setminus\mathcal{D}_0$, an LLM assigns a label $y_j \in \mathcal{T}_m \cup \{\texttt{other}\}$, where $\mathcal{T}_m$ is the taxonomy after $m$ iterations. 

\noindent
\underline{\textbf{Step 3 (Refinement).}} Instances labeled as \texttt{other}, i.e., $\mathcal{O}_m = \{x_j \mid y_j=\texttt{other}\}$, are further inspected by human annotators. If new failure patterns are identified, new modes $\Delta \mathcal{T}$ are added to the taxonomy, yielding $\mathcal{T}_{m+1} = \mathcal{T}_m \cup \Delta \mathcal{T}$.

This loop repeats until $\Delta \mathcal{T}=\varnothing$, i.e., no new failure modes are found. We ultimately find the stabilized set of failure modes: $\mathcal{T}^* = \lim_{m \to \infty} \mathcal{T}_m$ (see Algorithm \ref{algo:rq2}).

Note that determining vulnerability types (failure modes) requires comparing failure cases with the ground-truth answers (or references). The specific methods tailored to each vulnerability type are detailed in Appendix \ref{app:decide-type}.

\subsection{Human-in-the-loop decision (RQ$_3$)}
\label{ssec:human-loop}

To balance reliability and scalability in large-scale categorization, we integrate human inspection with multi-agent LLM decisions:

\noindent
\underline{\textbf{Step 1 (Multi-agent decision).}} For each threat instance $x_i$ requiring failure mode classification, we construct a model set $\Theta = \{\text{GPT-5}, \text{Llama-4-17B}, \text{Gemini-2.5}, \text{Claude-Sonnet-4}\}$. Each model independently proposes an initial label $\hat{y}_i^{(1)} \in \mathcal{T}$ during the first execution.

\noindent
\underline{\textbf{Step 2 (Repetition for stability).}} Motivated by evidence that unstable internal knowledge may lead to fluctuations across runs \citep{kumar2024training}, we execute a second round of multi-agent deliberation. In this round, each LLM observes the predictions $\hat{y}_i^{(1)}$ from other models, refines its reasoning, and then proposes a new label $\hat{y}_i^{(2)}$ for the same instance.

\noindent
\underline{\textbf{Step 3 (Human verification).}} Human inspectors evaluate consistency along two dimensions: (1) if agent-level votes show disagreement (lack of consensus), or (2) if $\hat{y}_i^{(1)} \neq \hat{y}_i^{(2)}$ (fluctuation across rounds). In either case, the instance is flagged as uncertain:
$$
\mathcal{U} = \{x_i \mid \text{Var}(\{\hat{y_i}^{(a)}\}) > 0 \;\;\text{or}\;\; \hat{y}_i^{(1)} \neq \hat{y}_i^{(2)} \}
\label{eq:uncertain}
$$
where $\{\hat{y_i}^{(a)}\}$ denotes the set of labels assigned by agents $a \in \Theta$. All $x_i \in \mathcal{U}$ are then inspected by human annotators for final determination.

The multi-agent collaboration efficiently handles most of routine cases , while humans focus only on instances with instability or misalignment (empirically, less than 1.8\% cases in this step), thus balancing scalability with reliability in failure mode categorization (Algorithm in \ref{algo:rq3}).

\section{Experimental Result}
\label{sec:expt}

Our categorization method leads to a detailed analysis of vulnerability types, as presented in Figure \ref{fig:summary}. Building on this, we further investigate specific failure cases and potential defense strategies to address the following questions:

\noindent
\textbf{Q$_1$:} Is the proposed method effective in categorizing vulnerabilities within the threat landscape? 

\noindent
\textbf{Q$_2$:} How do different components of the proposed method influence its overall effectiveness? \

\noindent
\textbf{Q$_3$:} How do different LLMs exhibit varying categorizations? 

\noindent
\textbf{Q$_4$:} What are the root causes of these vulnerabilities? 

Next, we conduct detailed experiments to answer these questions using the same settings and models as in \S\ref{ssec:eval}. We further detail our defensive solutions in \S\ref{sec:defense}.

\begin{table*}[t]
  \footnotesize
  \centering
  \caption{Distribution (ratio) of categorized CTI vulnerabilities across different methods. \textbf{Given the absence of ground truth (hard labels), human annotation serves as a near-gold standard. Performance closely matching human annotation demonstrates a near-human capability for vulnerability discovery.}}
  \label{tab:baseline}
  \setlength{\tabcolsep}{2pt} 
  \resizebox{\textwidth}{!}{
  \begin{tabular}{l|ccccc|cccccc|cccc}
    \toprule
    \multirow{4.5}{*}{\bf Method} & \multicolumn{5}{c|}{\bf Spurious Correlation} & \multicolumn{6}{c|}{\bf Contradictory Knowledge} & \multicolumn{4}{c}{\bf Constrained Generalization} \\
    \cmidrule(lr){2-6} \cmidrule(lr){7-12} \cmidrule(lr){13-16}
     & \bf 1.1 & \bf 1.2 & \bf 1.3 & \bf 1.4 & \bf 1.5 & \bf 2.1 & \bf 2.2 & \bf 2.3 & \bf 2.4 & \bf 2.5 & \bf 2.6 & \bf 3.1 & \bf 3.2 & \bf 3.3 & \bf 3.4 \\
     & Co-men. & Exploit. & Confond.  & Skewed & Hier. & Temporal & Conflict  & Semantic & Diver. & Misal. & Counter. & Distribut. & Unseen & Overfit. & Env. \\
     & bias & bias & factor &  source & metadata & contra. & report & conflict & data & knowl. & CTI & bias & pattern & reasoning & unaware \\
    \midrule
     Human annotation & 4.14 & 1.71 & 15.72 & 6.16 & 2.05 & 13.98 & 22.29 & 4.14 & 4.53 & 1.97 & 7.45 & 21.79 & 23.13 & 7.64 & 11.76 \\
    \midrule
    Clustering & 8.53 & 5.24 & 10.16 & 12.41 & 6.83 & 18.37 & 15.68 & 10.21 & 9.87 & 6.54 & 14.26 & 18.59 & 16.42 & 15.84 & 16.23 \\
    Topic modeling & 2.19 & 0.57 & 22.32 & 3.14 & 0.42 & 8.43 & 30.12 & 1.28 & 2.34 & 0.69 & 3.52 & 32.41 & 12.57 & 2.16 & 4.89 \\
    GPT judge  & 10.27 & 7.13 & 12.59 & 8.92 & 5.37 & 10.51 & 19.84 & 8.46 & 8.12 & 5.58 & 11.23 & 20.17 & 18.29 & 14.38 & 13.94 \\
    Gemini judge  & 5.82 & 3.42 & 14.23 & 7.56 & 3.29 & 15.19 & 21.07 & 5.53 & 5.86 & 3.13 & 9.47 & 21.54 & 26.83 & 11.27 & 11.62 \\
    Multi-agent debate  & 3.91 & 4.18 & 16.87 & 5.29 & 4.81 & 12.24 & 24.53 & 6.89 & 3.27 & 4.62 & 6.18 & 25.46 & 20.61 & 12.42 & 8.57 \\
     \midrule
     \rowcolor{cellshade}
    \bf Ours   & 4.12 & 1.87 & 15.50 & 6.31 & 1.98 & 13.86 & 22.33 & 3.96 & 4.49 & 1.85 & 7.77 & 23.34 & 25.06 & 10.02 & 10.80 \\
    \bottomrule
  \end{tabular}}
\end{table*}

\begin{table*}[t]
  \footnotesize
  \centering
  \caption{Ablation study by removing components from proposed vulnerability categorization method. Performance closely matching human annotation demonstrates a near-human capability for vulnerability discovery.}
  \label{tab:ablation}
  \setlength{\tabcolsep}{2pt} 
  \resizebox{\textwidth}{!}{
  \begin{tabular}{l|ccccc|cccccc|cccc}
    \toprule
    \multirow{4.5}{*}{\bf Method} & \multicolumn{5}{c|}{\bf Spurious Correlation} & \multicolumn{6}{c|}{\bf Contradictory Knowledge} & \multicolumn{4}{c}{\bf Constrained Generalization} \\
    \cmidrule(lr){2-6} \cmidrule(lr){7-12} \cmidrule(lr){13-16}
     & \bf 1.1 & \bf 1.2 & \bf 1.3 & \bf 1.4 & \bf 1.5 & \bf 2.1 & \bf 2.2 & \bf 2.3 & \bf 2.4 & \bf 2.5 & \bf 2.6 & \bf 3.1 & \bf 3.2 & \bf 3.3 & \bf 3.4 \\
     & Co-men. & Exploit. & Confond.  & Skewed & Hier. & Temporal & Conflict  & Semantic & Diver. & Misal. & Counter. & Distribut. & Unseen & Overfit. & Env. \\
     & bias & bias & factor &  source & metadata & contra. & report & conflict & data & knowl. & CTI & bias & pattern & reasoning & unaware \\
    \midrule
     Human annotation & 4.14 & 1.71 & 15.72 & 6.16 & 2.05 & 13.98 & 22.29 & 4.14 & 4.53 & 1.97 & 7.45 & 21.79 & 23.13 & 7.64 & 11.76 \\
    \midrule
    w/o stratification & 12.62 & 5.14 & 10.37 & 11.23 & 5.89 & 17.45 & 16.12 & 9.87 & 8.45 & 5.23 & 12.56 & 17.89 & 16.34 & 14.56 & 15.23 \\
    w/o human initiation & 4.93 & 2.56 & 14.23 & 7.82 & 3.12 & 12.45 & 20.56 & 5.45 & 5.78 & 2.89 & 9.12 & 20.34 & 21.67 & 11.45 & 12.34 \\
    w/o multi-agent decision & 2.34 & 6.78 & 21.56 & 3.45 & 0.56 & 9.12 & 28.45 & 1.67 & 2.89 & 0.92 & 4.12 & 30.23 & 13.45 & 5.67 & 6.78 \\
    w/o repetition  & 9.45 & 6.23 & 11.67 & 10.12 & 6.34 & 19.56 & 17.89 & 9.12 & 7.56 & 6.12 & 13.45 & 19.67 & 17.23 & 15.45 & 14.56 \\
     w/o human verification  & 3.67 & 2.89 & 16.45 & 5.45 & 2.67 & 14.89 & 23.67 & 4.89 & 3.78 & 2.56 & 6.45 & 23.12 & 24.56 & 9.34 & 10.23 \\
    \midrule
    \rowcolor{cellshade}
    \bf Ours   & 4.12 & 1.87 & 15.50 & 6.31 & 1.98 & 13.86 & 22.33 & 3.96 & 4.49 & 1.85 & 7.77 & 23.34 & 25.06 & 10.02 & 10.80 \\
    \bottomrule
  \end{tabular}}
\end{table*}

\subsection{Effectiveness of vulnerability categorization with baselines (Q$_1$)}

{\bf Baseline setting.} To validate the efficacy of our proposed framework, we benchmark it against several basic methodologies: \textbf{(i) Clustering} aggregates evaluation results of CTI benchmarks (\S\ref{ssec:eval}) into unsupervised groups based on vector embedding similarity using \textit{K-Means}. \textbf{(ii) Topic modeling} leverages latent dirichlet allocation to statistically extract semantic structures and keywords from the benchmarking results (\S\ref{ssec:eval}). \textbf{(iii) GPT judge}, wherein GPT-5 independently classifies failure modes based on provided definitions. \textbf{(iv) Gemini judge} similarly leverages the Gemini model to categorize vulnerabilities. \textbf{(v) Multi-agent debate} constructs an iterative consensus mechanism where multiple LLM agents (same as in \S\ref{ssec:human-loop})  refine each other’s classifications to reduce individual hallucinations.

{\bf Results.} The comparison in Table \ref{tab:baseline} shows clear distributional gaps between unsupervised baselines and semantic reasoning approaches. Statistical methods such as clustering and topic modeling miss many of the fine-grained subtype boundaries, while producing a skewed failure distribution that departs substantially from the human reference. LLM-based judges align more closely with human judgments, but still exhibit drift on the more challenging categories that require multi-step reasoning. Overall, our method yields a subtype distribution that is consistently closest to human annotation, correcting the systematic skews seen in both single-agent predictions and simple consensus-style debates.

{\bf Explanations.} The remaining gap largely reflects how well each method handles context-dependent ambiguity. Topic modeling, for example, struggles with \textit{Confounding factors (1.3)} and \textit{Temporal contradictions (2.1)} because these errors are driven by latent causal structure and logical inconsistency rather than surface-level lexical similarity. Standard LLM judges reduce this mismatch, but they often over-attribute cases to \textit{Spurious correlations (1.x)} and inflate \textit{Overfitted reasoning (3.3)}, effectively “finding” patterns that are plausible in the abstract yet unsupported by the specific incident context. Our approach mitigates these tendencies by grounding decisions in explicitly retrieved evidence and structured multi-agent deliberation, which improves attribution for high-nuance categories such as \textit{Conflicting reports (2.2)} and \textit{Unseen patterns (3.2)} that other methods frequently misclassify.

\begin{figure*}[t]
  \centering
  \includegraphics[width=0.95\linewidth]{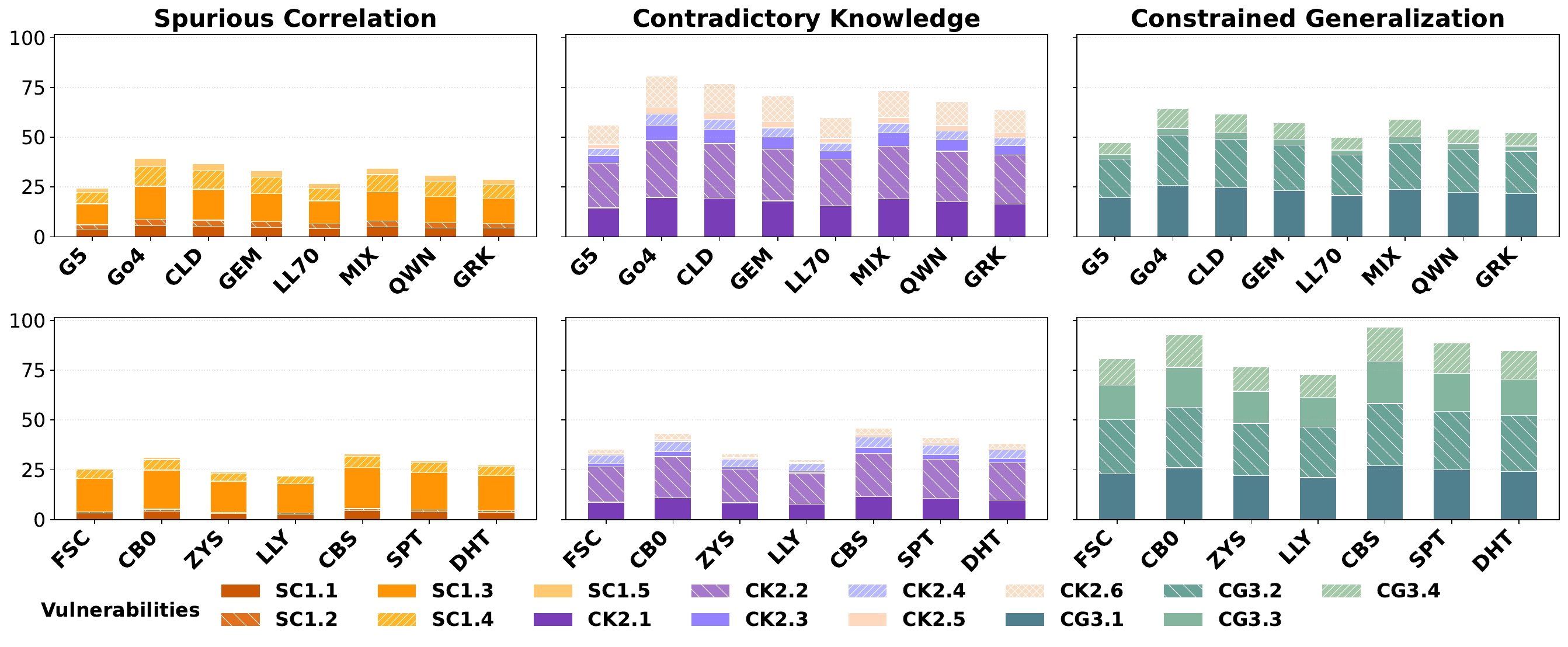}
  \caption{Varying proportions of vulnerabilities (types listed in Figure \ref{fig:summary}). Note that different vulnerabilities can intertwine within the same instance, which is particularly common in Contradictory Knowledge (CK) and Constrained Generalization (CG), less common in Spurious Correlation (SC).}
  \label{fig:failure_modes_specialized}
\end{figure*}

\begin{table*}[!t]
\small
\def\arraystretch{1}
\setlength{\tabcolsep}{4pt}
\caption{Summarized root causes of vulnerabilities, with detailed analysis in \S\ref{ssec:root-cause} and Appendix \ref{app:more-root-cause}.}
\centering
\resizebox{\textwidth}{!}{%
\begin{tabular}{lllp{14cm}}
\toprule
\textbf{Vulnerability} & \textbf{Subtype} & \textbf{Stage} & \textbf{How Vulnerability Happens} \\
\midrule
Spurious & (1.1) Co-mention bias & \ding{182}\ding{183}\ding{184}\ding{185}  & RAG surfaces unrelated but co-mentioned mitigations or vulnerabilities, causing LLMs to infer false associations. \\
 Correlation & (1.2) Exploitation bias & \ding{182}\ding{183}\ding{184}\ding{185} & Reused IoCs across incidents mislead models into over-attributing infrastructure reuse or ongoing activity. \\
 & (1.3) Confounding factors & \ding{182}\ding{183}\ding{184}\ding{185} & Non-causal variables (e.g., geography, org tags) treated as causal predictors for exploit likelihood. \\
 & (1.4) Skewed source & \ding{182} & Overrepresentation of certain feeds biases patch prioritization and defensive rules toward specific vendors. \\
 & (1.5) Hierarchical metadata & \ding{182}\ding{183}\ding{184} & Structured taxonomies (e.g., ATT\&CK chains) interpreted as causal orderings rather than descriptive metadata. \\
\midrule
 Contradictory & (2.1) Temporal contradiction & \ding{182}\ding{183}\ding{184}\ding{185} & Outdated advisories conflict with newer reports, confusing model reasoning about valid mitigations. \\
 Knowledge & (2.2) Conflicting reports & \ding{182}\ding{183}\ding{184}\ding{185} & Disagreement between sources on documented attack context, actor, dependencies, or other evidence. \\
 & (2.3) Semantic conflict & \ding{182} & Different naming/taxonomies (e.g., PlugX vs. Korplug) cause inconsistency.  \\
 & (2.4) Divergent structures & \ding{182}\ding{183}\ding{184} & JSON feeds vs.\ unstructured PDFs produce inconsistencies when fused into CTI mapping (e.g., TTP to patch). \\
 & (2.5)  Misaligned standards & \ding{183}\ding{184}\ding{185} & Differences in scoring frameworks (CVSS vs.\ vendor ratings) yield contradictory threat intelligence. \\
 & (2.6) Counteracting generation & \ding{183}\ding{184}\ding{185} & Reasoning and generation on CTI tasks disrupt LLM safety alignment, causing unstable outputs. \\
\midrule
Constrained
 & (3.1) Distributional bias & \ding{182} & Training on limited language/region corpora hinders generalization to unseen threat contexts. \\
Generalization & (3.2) Unseen patterns  & \ding{183}\ding{184}\ding{185} & Zero-day exploits exhibit novel paths absent from model training, degrading exploitability forecasts. \\
 & (3.3) Overfitted reasoning  & \ding{183}\ding{184} & Memorized patterns (e.g., CVE-TTP) lead to brittle linking and ineffective generation. \\
 & (3.4) Environmental unawareness  & \ding{183}\ding{184}\ding{185} & Models overlook local system/sector-specific dependencies, producing ineffective mitigation strategies. \\
\bottomrule
\end{tabular}
}
\label{tab:root-causes} 
\end{table*}

\subsection{Ablation study (Q$_2$)}

{\bf Results.} The ablation study in Table \ref{tab:ablation} shows a clear dependency across components. When we remove core architectural modules, especially multi-agent debate and stratification, the predicted subtype distribution departs sharply from the human gold standard, with unstable spikes emerging in categories such as \textit{Confounding Factors (1.3)} and \textit{Overfitted Reasoning (3.3)}. By comparison, disabling the human-in-the-loop steps (human initiation and verification) causes a smaller but still noticeable drop: the distribution drifts away from the ground truth, yet it remains far more stable than in settings where the automated reasoning machinery is removed.

\textbf{Explanations and efficiency analysis.} These results highlight a primary strength of our framework: \textbf{it cuts manual workload substantially without sacrificing efficacy.} Notably, even the w/o human verification variant still outperforms fully automated baselines (e.g., Clustering and Topic Modeling), indicating that the multi-agent architecture is doing most of the cognitive heavy lifting for failure categorization. In practice, this reframes the analyst’s role from labor-intensive labeling to lightweight oversight—spot-checking and validating outputs rather than building labels from scratch. As a result, CTI teams can approach near-human precision while spending only a fraction of the time typically required for manual classification.

\subsection{Varying LLM performance (Q$_3$)}

{\bf Results and explanations.} We further study the varying performance of different models. As shown in Figure \ref{fig:failure_modes_specialized}, and consistent with the distributions across vulnerabilities in Figure \ref{fig:summary}, our analysis reveals clear differences in how various LLMs handle distinct categories of errors: For vulnerabilities that directly impair LLM inference (e.g., co-mentioned but irrelevant mitigation strategies or outdated metadata that reduce retrieval effectiveness), general-purpose LLMs tend to accumulate a higher volume of failures. This indicates that their broad but non-specialized training leaves them vulnerable to overfitting and misinterpreting contextual evidence.

In contrast, for vulnerabilities rooted in the fine-tuning data (e.g., confounding factors in the threat corpus), cyber agents such as Foundation-Sec or Cyber-Zero tend to produce less reliable outputs (e.g., forecasting exploitability with contradictory PoCs). Such vulnerabilities act as data poisoning, where spurious correlations and contradictory knowledge can arise either intentionally (introduced by adversaries) or unintentionally (inherent in the fragmented cyber threat landscape).

Besides, we also observed that all models show constrained performance when confronted with emerging or zero-day threats. Failure ratios are particularly high in specialized cyber agents, whose localized nature and narrower pre-trained knowledge bases limit their adaptability. Once their training cutoffs are reached, they lack the generalization capacity to extrapolate effectively to novel threats.

\subsection{Root cause analysis (Q$_4$)}
\label{ssec:root-cause}

We conduct vulnerability-specific inspections across different CTI stages, investigating why a vulnerability may arise in a given stage (or why it may not) (\S\ref{sssec:root-cause-single} and App. ~\ref{app:more-root-cause}). Besides, we observe that vulnerabilities can be intertwined within the same CTI stage; we provide a detailed analysis in \S\ref{sssec:root-cause-multi} and App. ~\ref{app:more-intertwin}.

\subsubsection{Analyzing vulnerabilities within a single CTI stage}
\label{sssec:root-cause-single}

\begin{figure*}[t] 
    \begin{casebox}[width=\textwidth]{Case Study I: co-mention bias (Type 1.1) in \ding{182}  contextualization}
    
    Microsoft Patch Tuesday advisories often list many CVEs under one product (e.g.,\textit{ Windows Server}) \citep{microsoft_patch_tuesday}. The contextualization pipeline ingests the bulletin and a threat report cites \textit{CVE-2021-34527 (PrintNightmare)}, GPT-5 incorrectly infers that all co-mentioned CVEs were exploited in the same campaign, creating false links between irrelevant CVEs and active threats.
    \end{casebox}

    \begin{casebox}[width=\textwidth]{Case Study II: conflicting report (Type 2.2) in \ding{185}  mitigation}
    In the 2021 \textit{Microsoft Exchange “Hafnium”} case \citep{pitney2022systematic}, some advisories claimed patches fully mitigated the threat, while others warned of persistent web shells post-patch. The conflicting reports gave Qwen retriever false assurance about their mitigation effectiveness.
    \end{casebox}

    \begin{casebox}[width=\textwidth]{Case Study III: The intertwined threats -- MOVEit (CVE-2023-34362)}
    Early CTI reporting showed \textbf{contextualization failure (Type 1.3)}: triage lumped the SQL injection into generic IIS exploitation, and \textit{co-mention bias} hid the specific bug in \texttt{moveitisapi.dll}. Later, \textbf{Attribution failure (Type 2.1)} misattributed the activity to \textit{Silence} due to the "Truebot" downloader, despite later evidence pointing solely to \textit{Cl0p (Lace Tempest)}. Finally, \textbf{Prediction failure (Type 3.2)} underestimated the \textit{Impact Forecast}, missing the supply-chain cascade (e.g., Zellis, BBC) because the model overgeneralized from typical single-victim ransomware cases.
    \end{casebox}
\end{figure*}

During failure categorization (\S\ref{sec:method}), we gained insight into how different vulnerabilities are triggered by the techniques used across CTI tasks. Table~\ref{tab:root-causes} summarizes these drivers, aligned with the vulnerability types in Figure~\ref{fig:summary}. Below, we highlight representative cases.

\textbf{Co-mention bias (Type 1.1) in \ding{182} contextualization.}
As shown in Case Study I, co-mention bias arises during contextualization when retrieval systems treat entities that co-occur in raw reports as if they were causally related. Security bulletins often list multiple vulnerabilities, indicators, or mitigations in a single section, even though only a subset is relevant to the incident at hand. Methods such as topic modeling or coarse knowledge-base mapping, which lack fine-grained disambiguation, may surface all co-mentioned entities as equally relevant. This bloats the retrieved context with spurious links (e.g., attaching unrelated CVEs or MITRE TTPs to the same intrusion), which in turn prompts LLMs to propagate incorrect associations.

\textbf{Conflicting reports (Type 2.2) in \ding{185} mitigation.}
As shown in Case Study II, mitigation guidance often diverges across vendors due to differences in testing environments, threat visibility, and evaluation scope. For example, one advisory may claim a patch fully addresses an exploited vulnerability, while another provides evidence of continued exploitation via chained dependencies (e.g., an adjacent misconfiguration). Mitigation mapping built on such contradictory sources can overestimate or underestimate the effectiveness of a countermeasure. Similarly, summarization pipelines may produce inconsistent playbooks (e.g., some prioritizing patch deployment and others emphasizing compensating controls) leaving LLMs uncertain about actions truly mitigate the incident.

\textbf{Semantic conflict (Type 2.3) in \ding{183} attribution (less influential).}
Semantic conflict is less influential in attribution because actor identification depends on relational evidence and behavioral signatures. In practice, attribution relies more on infrastructure reuse, campaign TTPs, and stylistic cues than on whether a malware family is labeled ``PlugX'' or ``Korplug.'' Relation extraction and event-graph construction emphasize these structural patterns, making attribution models relatively tolerant to naming inconsistencies as long as the underlying relationships remain stable. While semantic divergence can introduce minor noise, it rarely flips the actor assignment: different labels for the same family typically still connect to the same infrastructure nodes and TTPs. As a result, attribution is comparatively robust to semantic conflicts.

A comprehensive analysis with case studies across all vulnerability types is provided in Appendix~\ref{app:more-root-cause}.

\subsubsection{Analyzing intertwined vulnerabilities}
\label{sssec:root-cause-multi}

As shown in Case Study III, we also observe that multiple vulnerabilities may interact within the same CTI instance. This stems from the tightly coupled nature of CTI pipelines: early-stage retrieval issues (e.g., co-mention bias or source skew) are carried forward as ``facts,'' leaving downstream attribution and prediction models with little room to falsify them. The problem is further amplified by heterogeneous and drifting evidence that force LLMs to memorize signals that are not truly compatible. Under this uncertainty, models fall back on inductive shortcuts (e.g., memorized actor-TTP associations), turning partial or noisy metadata into brittle reasoning. Finally, because CTI reasoning is hierarchical and dependency-driven (e.g., attack chains and mitigation mappings), misreading one link can cascade into adjacent stages. As a result, vulnerability combinations such as co-mention bias paired with temporal contradiction, or unseen patterns coupled with environmental unawareness, can form self-reinforcing loops that entangle multiple failures within a single instance.

Appendix~\ref{app:more-intertwin} provides additional analyses and case studies to complement this discussion.


\section{Potential Defensive Strategy}
\label{sec:defense}

\begin{table*}[!t]
  \footnotesize
  \def\arraystretch{0.85}
  \setlength{\tabcolsep}{5pt}
  \caption{Absolute improvement ($\Delta$) in CTI task performance following causal interventions. The proposed method targets all three failure types (spurious correlation, contradictory knowledge, and constrained Generalization) at inference time reasoning. Each number denotes the raw gain (e.g., F1 + $\Delta$) over the original performance (Table \ref{tab:eval}).}
  \label{tab:causal-improvement}
  \centering
  \resizebox{\textwidth}{!}{%
  \begin{tabular}{ll|cccccccc|ccccccc}
    \toprule
    \multirow{2.5}{*}{\bf Detailed CTI Task} & \multirow{2.5}{*}{\bf Metric} & \multicolumn{8}{c}{\bf Industry-leading, general-purpose} & \multicolumn{7}{c}{\bf Cybersecurity-specialized} \\
    \cmidrule(lr){3-10}\cmidrule(lr){11-17}
     & & G5 & Go4 & CLD & GEM & LL70 & MIX & QWN & GRK & FSC & CB0 & ZYS & LLY & CBS & SPT & DHT \\
    \lightmidrule
    \multicolumn{17}{c}{\bf \ding{182} Contextualization} \\
    \lightmidrule
    Affected Systems           & F1   & .031 & .042 & .055 & .068 & .022 & .061 & .054 & .041 & .038 & .041 & .035 & .033 & .039 & .042 & .038 \\
    Attack Infrastructure      & F1   & .025 & .048 & .062 & .071 & .031 & .068 & .065 & .059 & .052 & .055 & .041 & .038 & .045 & .048 & .044 \\
    Vulnerability Linking      & Acc  & .042 & .051 & .068 & .075 & .048 & .072 & .068 & .062 & .058 & .061 & .039 & .035 & .041 & .044 & .039 \\
    Malware Family Mapping     & F1   & .045 & .058 & .071 & .082 & .055 & .079 & .074 & .068 & .062 & .065 & .048 & .042 & .051 & .055 & .049 \\
    IOC Normalization          & F1   & .018 & .025 & .035 & .041 & .022 & .038 & .035 & .031 & .029 & .032 & .021 & .018 & .025 & .028 & .022 \\
    Threat Report Alignment    & BLEU & .082 & .095 & .112 & .098 & .075 & .105 & .098 & .089 & .091 & .085 & .062 & .058 & .075 & .088 & .065 \\
    Event Timeline Construction& BLEU & .075 & .088 & .095 & .091 & .082 & .094 & .089 & .081 & .085 & .082 & .055 & .048 & .058 & .062 & .055 \\
    Graph Population           & Acc  & .035 & .052 & .075 & .081 & .042 & .085 & .072 & .088 & .068 & .075 & .045 & .042 & .038 & .065 & .035 \\
    Source Reliability Scoring & AUC  & .012 & .015 & .028 & .021 & .018 & .032 & .025 & .022 & .025 & .028 & .015 & .012 & .018 & .022 & .015 \\
    \lightmidrule
    \multicolumn{17}{c}{\bf \ding{183} Attribution} \\
    \lightmidrule
    Threat Actor Linking       & Acc  & .021 & .028 & .055 & .042 & .035 & .052 & .048 & .065 & .058 & .062 & .025 & .018 & .028 & .025 & .035 \\
    TTP Extraction             & F1   & .032 & .038 & .048 & .052 & .045 & .055 & .051 & .048 & .042 & .045 & .031 & .028 & .035 & .038 & .032 \\
    Campaign Attribution       & Acc  & .048 & .055 & .068 & .072 & .052 & .075 & .071 & .065 & .058 & .061 & .042 & .038 & .045 & .048 & .042 \\
    Infrastructure Reuse       & F1   & .052 & .061 & .075 & .081 & .058 & .082 & .078 & .072 & .065 & .068 & .045 & .041 & .048 & .052 & .045 \\
    Language/Style Profiling   & Acc  & .025 & .032 & .045 & .048 & .038 & .052 & .048 & .045 & .041 & .044 & .028 & .025 & .032 & .035 & .028 \\
    False Flag Detection       & F1   & .088 & .095 & .108 & .112 & .082 & .115 & .105 & .098 & .095 & .092 & .075 & .068 & .082 & .085 & .078 \\
    Evidence Weighting         & BLEU & .092 & .105 & .125 & .115 & .088 & .118 & .112 & .105 & .098 & .095 & .072 & .065 & .085 & .092 & .075 \\
    Relation Graph Building    & F1   & .042 & .051 & .062 & .068 & .048 & .065 & .061 & .058 & .052 & .055 & .038 & .035 & .042 & .045 & .038 \\
    \lightmidrule
    \multicolumn{17}{c}{\bf \ding{184} Prediction} \\
    \lightmidrule
    Exploit Likelihood         & AUC  & .015 & .022 & .035 & .038 & .018 & .042 & .038 & .032 & .045 & .028 & .022 & .018 & .015 & .025 & .018 \\
    Impact Forecast            & BLEU & .065 & .078 & .092 & .085 & .058 & .095 & .088 & .082 & .075 & .072 & .052 & .048 & .058 & .065 & .052 \\
    Target Sector Prediction   & Acc  & .028 & .035 & .052 & .045 & .032 & .055 & .051 & .058 & .048 & .042 & .035 & .031 & .028 & .038 & .032 \\
    Campaign Escalation        & AUC  & .045 & .058 & .072 & .078 & .052 & .075 & .071 & .068 & .062 & .065 & .048 & .045 & .052 & .055 & .048 \\
    \lightmidrule
    \multicolumn{17}{c}{\bf \ding{185} Mitigation} \\
    \lightmidrule
    Patch Recommendation       & F1   & .038 & .045 & .058 & .062 & .032 & .065 & .061 & .055 & .052 & .055 & .041 & .038 & .032 & .035 & .045 \\
    Rule Generation (YARA)     & BLEU & .052 & .065 & .082 & .075 & .058 & .085 & .078 & .072 & .068 & .065 & .045 & .042 & .052 & .055 & .048 \\
    Response Summarization     & BLEU & .042 & .055 & .068 & .062 & .048 & .072 & .065 & .058 & .055 & .052 & .038 & .035 & .032 & .045 & .038 \\
    Mitigation–TTP Mapping     & Acc  & .035 & .042 & .055 & .058 & .041 & .062 & .058 & .052 & .048 & .051 & .032 & .028 & .031 & .035 & .032 \\
    Defensive Playbook Gen     & BLEU & .062 & .075 & .088 & .082 & .068 & .092 & .085 & .078 & .072 & .068 & .055 & .051 & .052 & .058 & .055 \\
    Countermeasure Ranking     & NDCG & .055 & .068 & .082 & .078 & .061 & .085 & .078 & .072 & .065 & .068 & .048 & .045 & .042 & .055 & .058 \\
    Incident Ticket Generation & Acc  & .025 & .032 & .045 & .038 & .018 & .048 & .045 & .042 & .038 & .045 & .028 & .025 & .022 & .028 & .025 \\
    \bottomrule
  \end{tabular}
  }
\end{table*}

Based on the experimental insights (\S\ref{sec:expt}), we further counteract CTI vulnerabilities at two levels of granularity: (1) a universal intervention applied during inference-time LLM reasoning to defend against all potential vulnerabilities, and (2) a set of fine-grained, qualitative mitigations applied to cyberspace.

\subsection{A causal intervention design to validate and mitigate CTI vulnerabilities}

\textbf{Design principle: counterfactual consistency.} 
Despite significant differences, the three vulnerability classes (spurious correlation, contradictory knowledge, and constrained generalization) share a common failure pattern in LLM reasoning: models tend to overfit to surface-level statistical co-occurrences rather than the \emph{invariant causal attributes} that actually determine the correct CTI judgment.

\begin{greencolorbox}
\ul{\textbf{Definition \faPencilSquareO: Causal attributes}} refer to CTI factors that should {causally drive} the label or conclusion, e.g., the attacker’s core \textit{TTPs} (techniques, procedures, and behavioral signatures), \textit{exploitation mechanism}, \textit{malware or tool behaviors/capabilities}, and other stable indicators of the underlying attack process, rather than incidental context such as timestamps, victim industry, geography, or reporting-source artifacts.

\end{greencolorbox}
To mitigate CTI vulnerabilities, we propose an inference-time \textbf{counterfactual consistency check} ($C^3$) based on a simple principle: if a CTI judgment is truly driven by these causal attributes, it should remain stable when we perturb \emph{non-causal} contextual tokens (e.g., dates, sectors) that should not affect the conclusion. Formally, we model CTI analysis as a mapping $f(x)\rightarrow y$. We then define a set of targeted intervention operators $\mathcal{I}=\{I_{\text{spur}}, I_{\text{contra}}, I_{\text{gen}}\}$ that generate counterfactual variants $x'_k$ by editing the non-causal tokens associated with each failure type, while preserving the CTI context (i.e., the causal attributes) that is leveraged to determine $y$.

\textbf{Validation (diagnosis).} 
To validate the existence of latent vulnerabilities without ground truth labels, we utilize $C^3$ as a sensitivity-based detector. For a given input $x$ and a target prediction $y$, we compute a \textit{Consistency Score} $\Omega$:
\begin{equation}
\Omega(y | x) = \frac{1}{|\mathcal{K}|} \sum_{k=1}^{|\mathcal{K}|} \mathbb{1} \left( \text{argmax}_{y'} P(y' | I_k(x)) = y \right)
\end{equation}
where $\mathcal{K}$ is the manifold of generated counterfactuals. If $\Omega(y|x) < \tau$ (a stability threshold), the prediction is flagged as ``$\texttt{Vulnerable}$.'' For example, if changing the report's date causes the model to flip its attribution from an APT identifier to ``unknown,'' we validate the presence of a \textit{contradictory knowledge} failure (Type 2.1). This metric provides a quantifiable proxy for model robustness, allowing us to detect specific failure modes in real-time.

\textbf{Inference-time mitigation (correction).} 
 The $C^3$ framework goes beyond diagnosis and can serve as an active mitigation engine. Instead of simply relying on the greedy decoding of the original input, we compute the final output $\hat{y}$ via a weighted consensus across the counterfactual set:
\begin{equation}
\hat{y} = \text{argmax}_y \sum_{k} w_k \cdot P(y | x'_k)
\end{equation}
where $w_k$ penalizes paths with high entropy. This effectively filters out reasoning driven by spuriousness  (Type 1.x), contradictions  (Type 2.x), or overfitting (Type 3.x) by forcing LLM reasoning to converge on evidence that is invariant across contexts.

\textbf{Experimental results.} We empirically show the efficacy of this design by applying the $C^3$ intervention across our benchmark tasks. Table \ref{tab:causal-improvement} details the absolute performance gains ($\Delta$) observed when switching from standard inference to our causal mitigation. We observe a universal improvement, particularly in tasks sensitive to high-noise metadata. For instance, \textit{false flag detection} presents a significant improvement (up to $+0.115$ F1 for Mixtral), demonstrating that spurious correlation was indeed the primary bottleneck. Similarly, \textit{event timeline construction} improves significantly across all models (avg $\Delta \approx 0.08$), showing that $C^3$ successfully mitigates \textit{contradictory knowledge} conflicts. Notably, while specialized agentic models show smaller relative gains due to their higher baseline, the intervention still yields consistent improvements, proving that these causal vulnerabilities are inherent to the LLM architecture rather than just domain knowledge gaps.

\subsection{Fine-grained mitigation tailored to specific vulnerability subtypes}
\label{app:failure-mitigation}

Building on the causal intervention results above, we now design a set of fine-grained, actionable mitigation strategies tailored to each vulnerability subtype.

\subsubsection{Mitigating spurious correlation failures}
\label{sssec:mitigation-spurious}

Spurious correlation failures arise when LLMs overgeneralize from patterns that are statistically associated but not causally linked. These failures typically stem from co-occurrence signals, reused observables, or implicit biases embedded in CTI report structures and metadata. To mitigate such issues, we recommend a combination of {\bf (i) retrieval-time} evidence filtering and {\bf (ii) prompt-level} constraints, which incorporate causal reasoning and uncertainty modeling to reduce the reliance on misleading associations and encourage evidence-based reasoning. Below, we detail tailored designs:

\ul{\bf 1.1 Co-mention bias from raw threat incident.}
\textit{(i) Retrieval:} run entity resolution + role labeling, then keep only entities with \emph{incident-role evidence} (e.g., “operator,” “infrastructure owner,” “victim”) and down-rank “mentioned-in-passing” names. \textit{(ii) Inference:} require an \emph{evidence link} for every predicted actor/tool (quote/span + relation type), and add a simple rule: \emph{reject if the only support is co-mention}.

\ul{\bf 1.2 Exploitation bias from deliberately reused IoCs.}
\textit{(i) Retrieval:} maintain an IoC reuse profile (first/last seen, frequency, and how many different campaigns it appears in) and down-rank “high-reuse” IoCs unless they come with \emph{campaign-specific} behavioral clues (delivery method, exploit chain, payload traits). \textit{(ii) Inference:} force a two-step check: first ask “is this IoC discriminative here, or commonly reused?”, then only make attribution/recommendations if it is discriminative.

\ul{\bf 1.3 Confounding factors that explicitly/implicitly correlate entities.}
\textit{(i) Retrieval:} separate \emph{causal evidence} (behavior, exploit mechanism, toolmarks, infrastructure ownership) from \emph{incidental context} (sector, region, time), and limit how much context-only metadata can drive retrieval ranking. \textit{(ii) Inference:} add a counterfactual prompt: “would the conclusion change if we changed sector/region/date?” If yes, lower confidence or abstain, and ask for causal evidence.

\ul{\bf 1.4 Skewed source.}
\textit{(i) Retrieval:} enforce source diversity (e.g., at least $k$ different publishers) and de-duplicate near-identical reports; attach source + timestamp tags to each claim. \textit{(ii) Inference:} require multi-source support for high-confidence claims; otherwise summarize the disagreement (“A says X, B says Y”) and output a ranked hypothesis list with uncertainty.

\ul{\bf 1.5 Hierarchical metadata from attack chains.}
\textit{(i) Retrieval:} first treat kill-chain stages as \emph{soft} hints unless backed by artifacts (telemetry/logs); then attach a confidence note to each stage transition. \textit{(ii) Inference:} add a strict rule: do not infer later-stage actions unless there is explicit evidence for them. If the chain contains incomplete evidence, keep the answer partial on present metadata.

\begin{table*}[t]
  \footnotesize
  \centering
  \caption{Ratio of vulnerabilities (same as Figure \ref{fig:summary}) after applying tailored mitigations. Lower ratio is better.}
  \label{tab:miti}
  \setlength{\tabcolsep}{2pt} 
  \resizebox{\textwidth}{!}{
  \begin{tabular}{ll|ccccc|cccccc|cccc}
    \toprule
    \multicolumn{2}{c|}{\multirow{4.5}{*}{\bf Method}} & \multicolumn{5}{c|}{\bf Spurious Correlation} & \multicolumn{6}{c|}{\bf Contradictory Knowledge} & \multicolumn{4}{c}{\bf Constrained Generalization} \\
    \cmidrule(lr){3-7} \cmidrule(lr){8-13} \cmidrule(lr){14-17}
     & & \bf 1.1 & \bf 1.2 & \bf 1.3 & \bf 1.4 & \bf 1.5 & \bf 2.1 & \bf 2.2 & \bf 2.3 & \bf 2.4 & \bf 2.5 & \bf 2.6 & \bf 3.1 & \bf 3.2 & \bf 3.3 & \bf 3.4 \\
     & & Co-men. & Exploit. & Confond.  & Skewed & Hier. & Temporal & Conflict  & Semantic & Diver. & Misal. & Counter. & Distribut. & Unseen & Overfit. & Env. \\
     & & bias & bias & factor &  source & metadata & contra. & report & conflict & data & knowl. & CTI & bias & pattern & reasoning & unaware \\
    \midrule
  & Retrieval-only & 3.12 & 2.05 & 10.45 & 2.15 & 2.11 & 4.25 & 6.82 & 3.55 & 2.12 & 1.15 & 9.45 & 18.45 & 19.56 & 2.55 & 3.12 \\
  General  & Inference-only & 1.85 & 1.15 & 5.65 & 5.23 & 1.89 & 10.45 & 14.56 & 2.89 & 3.15 & 1.95 & 5.12 & 12.34 & 11.45 & 1.12 & 5.45 \\
  & Full-fledged & 1.05 & 0.65 & 3.12 & 1.85 & 0.95 & 2.55 & 4.15 & 1.25 & 1.55 & 0.65 & 3.25 & 9.56 & 7.85 & 0.45 & 2.15 \\
    \midrule
  & Retrieval-only & 2.56 & 0.72 & 13.45 & 1.12 & 0.55 & 2.15 & 5.12 & 1.15 & 1.95 & 0.45 & 1.89 & 19.56 & 24.12 & 16.45 & 6.89 \\
  Agent  & Inference-only & 1.25 & 0.35 & 7.85 & 3.45 & 0.48 & 6.89 & 10.45 & 0.85 & 3.12 & 0.65 & 0.85 & 13.45 & 14.56 & 8.45 & 11.25 \\
   & Full-fledged & 0.85 & 0.22 & 4.56 & 0.95 & 0.25 & 1.55 & 3.45 & 0.45 & 1.25 & 0.25 & 0.45 & 10.56 & 11.25 & 5.12 & 5.45 \\
     
    \bottomrule
  \end{tabular}}
\end{table*}

\subsubsection{{Mitigating contradictory knowledge failures}}
\label{sssec:mitigation-contradiction}

Contradictions occur when models encounter conflicting, inconsistent, or incompatible CTI between different threat databases or platforms. These contradictions can arise from differences in temporal scope, analyst interpretation, data formats, or semantic misalignment. To mitigate these failures, systems should incorporate {\bf (i) retrieval-based} temporal or source-aware annotation and {\bf (ii) inference-time} conflict detection and uncertainty justification. Detailed strategies are:

\ul{\bf 2.1 Temporal contradiction between outdated and recent evidence.}
\textit{(i) Retrieval:} time-aware reranking (prefer recent + corroborated sources), and package evidence in chronological order with clear $\texttt{as-of}$ time tags. \textit{(ii) Inference:} when claims conflict, prefer newer evidence unless it is clearly weaker; otherwise present both with timestamps and explain the reason of conflict in a self-reflective generation.

\ul{\bf 2.2 Conflicting reports of attack contexts or dependencies.}
\textit{(i) Retrieval:} cluster retrieved passages into consistent views (by source or by claim similarity) and feed them as grouped bundles rather than a mixed list. \textit{(ii) Inference:} require a short divergence summary (e.g., which vendor and information source claim what scenarios) and a decision policy: choose one with justification, or abstain and list what extra evidence would resolve it.

\ul{\bf 2.3 Semantic conflict.}
\textit{(i) Retrieval:} canonicalize entities (aliases, malware family names, tool versions) using an alias table/KB, and attach disambiguation notes when names are similar. \textit{(ii) Inference:} add a gate before merging claims: “same entity, alias, or related-but-different?” and do not merge unless it is truly the same; also explicitly check for negation (e.g., not attributed).

\ul{\bf 2.4 Divergent data structure from different platforms.}
\textit{(i) Retrieval:} normalize different schemas (e.g., STIX, vendor formats) into one simple intermediate record (typed fields or triples) before passing to the model. \textit{(ii) Inference:} feed the model the normalized record with field definitions, and instruct it to reason from field-specific meaning (prompt derived from schemas).

\ul{\bf 2.5 Misaligned knowledge and security standards.}
\textit{(i) Retrieval:} attach authoritative references during retrieval (valid MITRE ATT\&CK technique IDs, CVSS rubric snippets, NVD-style conventions) and validate candidate mappings early. \textit{(ii) Inference:} constrain outputs to valid options (pick from a provided list); if evidence does not support a clean mapping, output “unknown/insufficient evidence” with a brief rationale.

\ul{\bf 2.6 Counteracting CTI generation and LLM alignment.}
\textit{(i) Retrieval:} require every major claim to carry a grounding span (quote + source + time), and keep an audit trail of which evidence supported which claim. \textit{(ii) Inference:} enforce concise bullet claims with a citation to evidence, plus calibrated uncertainty (no confident wording without support).

\subsubsection{Mitigating constrained generalization failures}
\label{sssec:mitigation-generalization}

Generalization issues in cybersecurity often present as insufficient understanding on emerging threats or adversarial shifts in attack behavior \citep{nam2019understanding}. To mitigate these issues, systems should incorporate  {\bf (i) retrieval-based} data augmentation to collect auxiliary knowledge from external sources and {\bf (ii) inference-time}  generalization (rather than memorization) to emphasize task-specific reasoning:

\ul{\bf 3.1 Distributional bias.}
\textit{(i) Retrieval:} balance CVE exposures by reweighting retrieval results so long-tail actors/regions/tools are not drowned out; add simple quotas or diversity constraints by geography/tooling when possible. \textit{(ii) Inference:} explicitly warn against common threat actors (or infrastructure) and require the deliberation to cite causal indicators (behavior/exploit/toolmarks), not just region/sector stereotypes.

\ul{\bf 3.2 Unseen pattern from emerging threats.}
\textit{(i) Retrieval:} plug in fresh CTI feeds (recent reports, disclosures, malware analysis notes) with freshness filters and credibility scoring, so the context reflects post-training threats. \textit{(ii) Inference:} use a novelty-aware response: propose cautious hypotheses, clearly mark what is unknown, and list the top evidence signals that would confirm or refute each hypothesis.

\ul{\bf 3.3 Overfitted reasoning.}
\textit{(i) Retrieval:} diversify evidence so one shortcut (e.g., tool $\rightarrow$ actor) cannot dominate; when possible, retrieve counterevidence or alternatives. \textit{(ii) Inference:} add an evidence-sufficiency gate: if attribution is not explicitly supported, lower confidence or abstain; optionally run a small counterfactual check that swaps common shortcuts (actor/infra) to see if the conclusion collapses.

\ul{\bf 3.4 Environmental unawareness.}
\textit{(i) Retrieval:} treat environment details as primary inputs (including but not limited to asset type, deployment, sector constraints, region) and retrieve environment-matched guidance and prior cases. \textit{(ii) Inference:} require conditional reasoning: ``given environment E, risk/mitigation changes because…'', and tie incident response recommendations to environment-specific constraints.

\subsubsection{Experiment results on mitigation}

The experimental results in Table \ref{tab:miti} demonstrate that the above approaches yield consistent reduction in vulnerability rates across all categories, outperforming both ablation baselines (i.e., retrieval-only or inference-only mitigation). A comparative analysis of the ablations reveals distinct component strengths. Retrieval-only mitigation proves superior in resolving contradictory knowledge (Types 2.1, 2.2, 2.4) and environmental unawareness (Type 3.4). This suggests that failures driven by outdated information or source ambiguity are best addressed by grounding the model in retrieved, verifiable evidence (e.g., reducing General-LLM temporal contradiction from 17.53\% to 4.25\%). This improvement implies that external evidence serves as an auxiliary signal for time-sensitive or source-conflicting queries.

Conversely, Inference-only mitigation dominates in mitigating spurious correlations (Types 1.1, 1.3) and constrained generalization (Types 3.2, 3.3). For instance, in \textit{confounding factors} (1.3), the inference-based check reduces the failure rate for General LLMs to 5.65\% (vs. 10.45\% for retrieval), indicating that these errors stem from flawed internal reasoning rather than a lack of information. The full-fledged mitigation further synthesizes retrieval to cleanse the input context and causal inference to filter the reasoning process, thereby achieving the lowest universal failure ratios.
\section{Related Work}

\textbf{LLM-as-agent.}  
LLM-based agents have been explored across diverse domains such as education~\citep{tang2025llms,chu2025llm,neumann2024llm}, scientific discovery~\citep{shojaee2024llm,ren2025towards}, healthcare~\citep{goyal2024healai,zhou2024zodiac,more2026theramind,meng2026small,brens2026semantic}, and urban mobility modeling~\citep{zhou2024urban,wang2024large,calderon2025cognitive}. These applications highlight the ability of LLM agents to decompose complex tasks, integrate external tools, and generate executable outputs.   

\textbf{LLMs for cybersecurity.}  
LLMs are increasingly leveraged for both offensive and defensive cybersecurity tasks, owing to their strong natural language understanding and reasoning capabilities. In static analysis, LSAST augments traditional SAST tools with dynamic vulnerability knowledge~\citep{keltek2025lsast}, while hybrid systems use LLM-driven preprocessing and explanation to improve anomaly detection in IoT~\citep{ghimire2025enhancing}. Autonomous agents have demonstrated the ability to exploit one-day vulnerabilities~\citep{fang2024llm} and guide fuzzing across multi-hop library dependencies~\citep{zhou2024magneto}. Additional applications include enhancing intrusion detection~\citep{g2024harnessing}, supporting large-scale code review~\citep{sun2025bitsai,liu2025cylens}, and enabling malware tracking through dataset augmentation and semantic analysis~\citep{yu2024maltracker,liu2025benchmarking,tang2025polar,sheikh2026riskbridge}.  

\textbf{Vulnerabilities of LLMs.}  
Despite their strong capabilities, LLM-based agents exhibit critical vulnerabilities that undermine reliability in high-stakes domains. Recent work shows that malfunction amplification attacks can cascade small reasoning errors into severe misjudgments \citep{zhang2025breaking,tang2026value}. Other studies highlight vulnerabilities to data poisoning \citep{wang2025comprehensive,xi2023defending}, adversarial prompts \cite{debenedetti2024agentdojo}, and misalignment \citep{fang2025preemptive,liu2025data}, all raise concerns for deploying LLMs in high-stakes environments.

\section{Conclusion}


This work presents systematic and intensive studies of intrinsic vulnerabilities that constrain the effectiveness of LLMs in cyber threat intelligence. We combine large-scale benchmark evaluations with real-world CTI reports and uncover three fundamental failure modes (spurious correlations, contradictory knowledge, and constrained generalization) that persist across multiple CTI stages and workflows. Our autoregressive, human-in-the-loop methodology enables reliable categorization of failure instances and provides insights into how these limitations emerge and propagate. Correspondingly, we proposed tailored sets of mitigation strategies to address different subtypes of vulnerabilities. Further evaluations demonstrate the preliminary effectiveness of these mitigation solutions. Overall, we not only reveal blind spots in current cybersecurity-centric LLM reasoning but also chart a path toward more principled model adaptations and robust cyber agent design.


\bibliographystyle{ACM-Reference-Format}
\bibliography{ccs2026conference}

\appendix

\section{Complementary Details of CTI}
\label{app:cti-stage}

This appendix section provides additional details on how different CTI stages are conducted with the involvement of various techniques, complementing Section~\S\ref{ssec:cti_stages}.

\subsection{\ding{182} Contextualization}

The contextualization stage transforms raw, fragmented observations into structured and actionable intelligence. Below we elaborate on the specific techniques commonly employed in real world CTI operations, with a natural sequence in practice.

\begin{itemize}
    \setlength{\leftmargin}{-10pt}
    \item \textbf{Topic Modeling.} Topic modeling techniques such as Latent Dirichlet Allocation (LDA) \citep{blei2003lda}, Non-negative Matrix Factorization \citep{lee1999nmf}, and more recent neural topic models \citep{dieng2020topic} are applied to large corpora of unstructured CTI text, including threat reports, incident tickets, and log annotations. These models group documents or paragraphs into coherent themes, enabling analysts to identify clusters of related activity such as ``phishing campaigns leveraging Office macros'' or ``ransomware families exploiting VPN vulnerabilities.'' In practice, topic modeling supports early triage by prioritizing threat feeds that share thematic overlap with active campaigns in a given sector.

    \item \textbf{Event Extraction.} Event extraction identifies structured incidents—\textit{who did what, when, and how}—from unstructured logs or reports. Natural Language Processing (NLP) pipelines detect entities such as vulnerabilities (CVEs), indicators of compromise (IP addresses, file hashes), and TTPs (MITRE ATT\&CK techniques \citep{strom2018mitre}), then associate them with temporal markers and actor actions. In Security Operations Centers (SOCs), event extraction enables long-form reports to be converted into structured JSON objects or STIX bundles, which can be automatically ingested into SIEM platforms such as Splunk or Elastic. This reduces manual parsing effort and ensures consistent representation across heterogeneous data sources.

    \item \textbf{Knowledge Base Mapping.} Knowledge base mapping aligns extracted entities and events with standardized taxonomies such as CVE \citep{cve}, CWE \citep{cwe}, MITRE ATT\&CK \citep{strom2018mitre}, and CAPEC \citep{capec}. This process typically involves both exact matching (e.g., direct CVE ID resolution) and approximate entity linking (e.g., mapping the phrase ``remote PowerShell execution'' to ATT\&CK T1059.001). In CTI practice, knowledge base mapping is essential for interoperability: intelligence can be shared across organizations using a common language of identifiers, enabling correlation of local incidents with global adversary behaviors.
    
    \item \textbf{Information Retrieval.} Information retrieval systems allow analysts to ground their findings in relevant historical reports and external databases. Implementations include keyword- and embedding-based retrieval from proprietary CTI feeds (e.g., Recorded Future, Mandiant Advantage), open-source repositories (e.g., AlienVault OTX, AbuseIPDB), and structured vulnerability catalogs such as NVD \citep{nvd} and the CISA KEV catalog \citep{CISA_KEV_Catalog}. Retrieval pipelines often combine lightweight keyword filters for precision with dense embedding search for semantic coverage. Analysts use these systems to verify whether a newly observed domain has prior associations with known malware campaigns, or whether a vulnerability is actively being exploited.

    \item \textbf{Operational Integration.} In SOC environments, the above techniques are integrated into semi-automated pipelines. For example, a suspicious DNS query may trigger automated enrichment: (i) topic modeling identifies thematic overlap with phishing campaigns, (ii) event extraction links the query to specific malware families, (iii) knowledge base mapping ties the observed behavior to ATT\&CK techniques, and (iv) information retrieval retrieves prior cases of infrastructure reuse. The result is a structured incident summary that equips analysts with actionable context for attribution, prediction, and mitigation.
\end{itemize}

\subsection{\ding{183} Attribution}

Attribution in cyber threat intelligence involves identifying the adversary or campaign responsible for observed malicious activity. This process requires combining technical indicators with contextual and behavioral evidence. Below we describe the primary techniques used in real-world practice, along with how they are operationalized by analysts.
\begin{itemize}
    \item \textbf{Named Entity Recognition (NER).} NER systems are used to extract structured entities such as malware families, infrastructure elements, or actor names from unstructured reports and incident logs. For example, extracting references to \textit{APT28}, \textit{Mimikatz}, or \textit{QakBot} across heterogeneous feeds helps analysts consolidate threat narratives.

    \item \textbf{Relation and Event Extraction.} Beyond isolated entities, attribution requires uncovering relationships among adversary tactics, techniques, and infrastructure. Relation extraction techniques map, for instance, shared IP ranges between phishing campaigns or code reuse across malware variants, while event extraction captures sequences of actions such as exploitation followed by lateral movement \citep{husari2017ttpdrill}.

    \item \textbf{Infrastructure Correlation.} Adversaries frequently reuse or repurpose command-and-control (C2) infrastructure. CTI teams apply graph-based correlation to link domain registrations, TLS certificates, and hosting providers, enabling attribution of otherwise fragmented observations to known actor toolkits or campaigns \citep{stone2009your,antonakakis2017understanding}.

    \item \textbf{Stylistic and Linguistic Profiling.} Analysts also consider linguistic cues in adversary communications or malware code artifacts. Stylometry and compilation fingerprinting methods can identify patterns in variable naming, debugging strings, or grammar usage, which help to tie malware development back to particular groups \citep{rosenblum2008learning}.


    \item \textbf{Campaign Graph Construction.} To synthesize diverse evidence, CTI analysts build structured graphs of adversarial campaigns, linking entities, infrastructure, and TTPs across incidents. Graph construction enables propagation of attribution hypotheses and detection of actor evolution over time.
\end{itemize}

\subsection{\ding{184} Prediction}

Prediction in CTI involves forecasting adversarial actions, exploitation likelihood, and campaign evolution. Unlike contextualization and attribution, prediction requires reasoning under temporal uncertainty and incomplete information. In practice, analysts and automated systems deploy a range of data-driven and model-based techniques to anticipate threats:

\begin{itemize}
    \item \textbf{Historical correlation and trend analysis.} Analysts correlate prior incidents and intrusion campaigns to identify recurring attacker playbooks. For example, statistical methods on longitudinal CVE exploitation data help assess whether recently disclosed vulnerabilities follow exploitation trends of past families \citep{bilge2012before, allodi2017economic}.

    \item \textbf{Exploit prediction scoring.} Models such as the Exploit Prediction Scoring System (EPSS) estimate the probability that a vulnerability will be exploited within a given time window, using features such as vulnerability metadata, CVSS scores, and real-world exploit observations \citep{bullough2017predicting}.

    \item \textbf{Temporal modeling of campaign progression.} Recurrent neural networks and temporal point processes capture how campaigns unfold over time, modeling likely transitions from initial access to follow-on payloads such as ransomware.

    \item \textbf{Temporal forcasting.} LLMs predict which TTPs an actor is likely to employ next or what's impact. This supports proactive defense, such as generating detection rules for tactics not yet observed in the ongoing campaign \citep{sood2012targeted}.

    \item \textbf{Threat propagation simulation (synthesis).} Agent-based and epidemic-style models simulate the spread of malware or worms across interconnected systems, forecasting infection curves and propagation likelihood \citep{cheng2010modeling}.
\end{itemize}

\subsection{\ding{185} Mitigation}

Mitigation is the final stage of cyber threat intelligence (CTI), where enriched analysis and attribution results are translated into concrete defensive measures. Unlike contextualization or attribution, which primarily generate insights, mitigation requires actionable transformations that directly alter security posture. Below, we introduce the major classes of techniques commonly adopted in real-world CTI practice.

\begin{itemize}

    \item \textbf{Detection rule generation.} Security teams design, validate, and deploy detection rules in languages such as Sigma and YARA to capture specific threat behaviors. In practice, detection rules are tuned iteratively: analysts translate threat reports into rule signatures, test them against telemetry or sandbox logs, and refine them to reduce false positives while ensuring coverage of attacker tradecraft.

    \item \textbf{Mitigation efficacy evaluation.} Beyond applying countermeasures, analysts must assess their effectiveness. Approaches include red-teaming exercises, breach-and-attack simulation (BAS) platforms, and adversary emulation scenarios that replay known TTPs to test whether mitigations succeed in preventing, detecting, or containing malicious activity \citep{shostack2014threat}.

    \item \textbf{Response playbook recommendation.} Structured playbooks standardize incident response by encoding lessons from CTI. These include step-by-step containment and recovery actions tailored to adversary campaigns (e.g., disabling compromised accounts, isolating infected subnets). Orchestration tools such as SOAR platforms automate playbook execution, integrating CTI feeds into dynamic workflows \citep{mitnick2017art}.

    \item \textbf{Summarization.} Finally, mitigation intelligence is communicated through concise, contextualized reports for executives and IT operators. Summarization synthesizes prioritized vulnerabilities, mapped mitigations, and recommended workflows. This ensures decision makers understand trade-offs between operational impact and security gains, and supports cross-team coordination in enterprise-scale defense \citep{mavroeidis2017cyber}.
\end{itemize}

\section{Additional Details of CTI Experiment for \S\ref{sec:task}}

\subsection{Used Benchmark in Evaluation}
\label{app:benchmark}

Table \ref{tab:src-bench-stat} presents the original scales of the datasets used in our evaluation.  

\begin{table*}[h]
  \centering
  \caption{Used cybersecurity benchmarks in our evaluations. \label{tab:src-bench-stat}}
  \begin{tabular}{l l c c c}
    \toprule
    \textbf{Benchmark} & \textbf{Focus} & \textbf{\#Data} & \textbf{\#Task} & \textbf{\#Source} \\
    \midrule
    CTIBench \citep{alam2024ctibench} & Cyber Threat Intelligence & 5,610 & 5 & N/A \\
    SevenLLM-Bench \citep{ji2024sevenllm} & Report Analyzing & 92,701 & 28 & N/A \\
    SWE-Bench \citep{jimenez2023swe} & Bug fixing & 2,294 & 12 & 1 \\
    CYBERTEAM \citep{liu2025benchmarking} & Blue-team threat hunting & 452,293 & 30 & 23 \\
    \bottomrule
    
  \end{tabular}
\end{table*}

\subsection{Used Real-World Databases and Platforms in Evaluation}
\label{app:src-db}

In addition to benchmarks, we incorporate several real-world cybersecurity databases and intelligence platforms to ensure that our evaluation settings reflect practical CTI usage. Each database provides complementary coverage across the CTI stages (\S\ref{ssec:cti_stages}), and we detail both their scope and our methodology for leveraging their information.

\textbf{National Vulnerability Database (NVD).} 
The NVD \citep{nvd} serves as the canonical repository for software vulnerabilities and their CVSS severity scores. We utilize NVD entries to support all \ding{182}--\ding{185} tasks. Specifically, we map vulnerabilities in benchmark items to NVD records in order to standardize CVE identifiers, extract official CVSS vector strings, and obtain temporal metadata (publication and modification dates). These fields allow us to align system-environment observations with ground-truth vulnerability characteristics, and to test forecasting models that predict exploit likelihood (e.g., by contrasting NVD base scores against EPSS-derived estimates).

\textbf{Exploit Prediction Scoring System (EPSS).} 
EPSS \citep{jacobs2021exploit} provides probabilistic estimates of the likelihood that a given CVE will be exploited in the wild. We use EPSS scores directly for \ding{184} prediction tasks, both as a source of labels (ground-truth exploitation likelihood) and as a reference distribution to evaluate calibration of LLM-based forecasts. EPSS time-series updates also enable temporal correlation experiments, where we assess whether models capture shifts in exploitation likelihood following disclosure, patch release, or threat-actor reuse.

\textbf{MITRE ATT\&CK and CAPEC.} 
The ATT\&CK \citep{mitreattack} knowledge base encodes adversarial tactics, techniques, and procedures (TTPs) in a structured taxonomy. We map CTI tasks involving entity extraction, campaign attribution, and mitigation alignment to ATT\&CK entries. For instance, when evaluating \ding{183} attribution, extracted TTPs from LLM outputs are compared against ATT\&CK technique identifiers to assess correctness. We also leverage CAPEC (Common Attack Pattern Enumeration and Classification) \citep{capec} to validate abstract attack patterns referenced in benchmark items, particularly for mapping contextualized logs or IOCs to higher-level adversarial behaviors.

\textbf{MISP (Malware Information Sharing Platform).} 
MISP \citep{MISP} serves as a community-driven threat intelligence sharing platform, containing structured feeds of indicators of compromise (IOCs), malware samples, and infrastructure metadata. We use MISP to enrich contextualization tasks (\ding{182}) by grounding benchmark instances in realistic IOC–malware–actor relationships. For example, when a dataset item involves resolving a suspicious domain, we verify its presence in MISP feeds and align it to associated threat actors or malware families. This enrichment supports evaluation of LLMs’ ability to normalize IOCs and link them to campaigns.

\textbf{VirusTotal.} 
VirusTotal \citep{VirusTotal} aggregates antivirus detections and malware analysis results across a large corpus of submitted files, domains, and URLs. We leverage VirusTotal reports for \ding{182} contextualization and \ding{183} attribution tasks. In contextualization, VirusTotal tags (e.g., malware family labels, sandbox behavior summaries) serve as auxiliary ground-truth for tasks like malware family mapping. In attribution, we analyze infrastructure overlap by checking whether related domains or IPs have been co-reported in VirusTotal samples, allowing us to validate LLM predictions about infrastructure reuse.

\textbf{Open Threat Exchange (OTX).} 
AlienVault’s OTX \citep{OTX} provides community-curated threat pulses (collections of IOCs associated with specific campaigns or malware). We use OTX primarily for \ding{183} attribution and \ding{184} prediction: pulses give us labeled groupings of IOCs tied to campaigns, which we then cross-check against LLM-predicted campaign attributions. 

\textbf{Security Advisories and Vendor Bulletins.} 
Finally, vendor advisories (e.g., Microsoft, Cisco, Progress Software) and public CERT bulletins provide authoritative patching and mitigation recommendations. We incorporate these resources into \ding{185} mitigation tasks by aligning recommended countermeasures with benchmark items. For instance, in patch recommendation evaluation, the correct answer set is derived from vendor bulletins rather than from secondary threat reports. Similarly, YARA and Sigma rule examples are drawn from advisory-linked repositories, ensuring that response summarization tasks are grounded in practical remediation steps.

For consistency, we design a preprocessing pipeline that (i) normalizes identifiers (CVE, IOC, ATT\&CK TTPs) across databases, (ii) aligns timeframes so that prediction tasks respect disclosure/exploitation chronology, and (iii) constructs ground-truth mappings between observations and actor/campaign/mitigation entities. We are thus able to systematically evaluate LLMs across all CTI stages using both controlled benchmarks (Table~\ref{tab:bench}) and real-world ground-truth data.

\subsection{Task Description}
\label{app:task-desc}

\noindent\textbf{\ding{182} Contextualization.}

\noindent\textbf{Affected Systems (F1).} Binary decision per asset: is a listed host/application impacted by the described CVE/IOC set (yes/no). \emph{Example:} decide whether \texttt{Exchange 2019 CU12} is affected given a CVE vector and server build.

\noindent\textbf{Attack Infrastructure (F1).} Binary decision per indicator: determine whether an IP/domain/URL belongs to adversary C2 or delivery infrastructure. \emph{Example:} classify \texttt{cdn-upd[.]com} as campaign infra vs. benign CDN.

\noindent\textbf{Vulnerability Linking (Acc).} Multi-class assignment of correct CVE(s) from candidates based on logs/snippets. \emph{Example:} map an IIS error pattern to \{\texttt{CVE-2021-34473}\} among distractors.

\noindent\textbf{Malware Family Mapping (F1).} Binary decision per candidate family: does observed behavior/artifacts match the family’s signature (yes/no). \emph{Example:} tag samples as belonging to a loader vs. banking trojan family.

\noindent\textbf{IOC Normalization (F1).} Binary correctness for canonicalizing raw IOCs (type+value) against gold forms. \emph{Example:} normalize \texttt{hxxp://ex[.]ample[.]com/login} to \texttt{http://ex.ample.com/login} (URL).

\noindent\textbf{Threat Report Alignment (BLEU).} Text similarity between a generated one-sentence alignment/abstract and a reference summary of the most relevant report. \emph{Example:} produce a synopsis that matches the gold advisory linkage.

\noindent\textbf{Event Timeline Construction (BLEU).} Compare generated event sequence text to a gold timeline. \emph{Example:} ``phish \textrightarrow\ beacon \textrightarrow\ lateral \textrightarrow\ exfil'' vs. reference steps.

\noindent\textbf{Graph Population (Acc).} Multi-label slot filling for nodes/edges in an event graph (accuracy over required triples). \emph{Example:} add \{host–used\_tool–T1059\} and \{user–compromised\_via–phish\} edges correctly.

\noindent\textbf{Source Reliability Scoring (AUC).} Binary scoring of source credibility (reliable vs. suspect) with probabilistic output; evaluated by ROC-AUC. \emph{Example:} score a paste site vs. vendor advisory on the same IOC claim.

\noindent\textbf{\ding{183} Attribution.}

\noindent\textbf{Threat Actor Linking (Acc).} Multi-class assignment of the most plausible actor profile(s) from candidates. \emph{Example:} pick the actor whose historical TTP set matches observed techniques.

\noindent\textbf{TTP Extraction (F1).} Binary decision per candidate technique ID: is T\#\#\#\# evidenced (yes/no). \emph{Example:} confirm \texttt{T1059} (command execution) from process tree snippets.

\noindent\textbf{Campaign Attribution (Acc).} Multi-class selection of a campaign label among candidates. \emph{Example:} assign activity to a 2023 spearphishing campaign vs. a 2024 credential-harvest run.

\noindent\textbf{Infrastructure Reuse (F1).} Binary decision per linkage: does an IOC show reuse across events (yes/no). \emph{Example:} mark \texttt{203.0.113.7} as reused across two clusters within 30 days.

\noindent\textbf{Language/Style Profiling (Acc).} Multi-class style attribution (e.g., build system, macro style, lure phrasing). \emph{Example:} assign documents to a known lure/style family.

\noindent\textbf{False Flag Detection (F1).} Binary decision: is an observed signature intentionally misleading (yes/no). \emph{Example:} detect planted strings mimicking a different actor’s toolkit.

\noindent\textbf{Evidence Weighting (BLEU).} Textual rationale summarizing which evidence most supports the conclusion; compared to a gold rationale via BLEU. \emph{Example:} generate a short justification prioritizing sandbox logs over OSINT.

\noindent\textbf{Relation Graph Building (F1).} Binary decision per candidate relation triple (entity–relation–entity). \emph{Example:} validate \{domain\(\rightarrow\)hosts\(\rightarrow\)IP\} and reject spurious actor edges.

\noindent\textbf{\ding{184} Prediction.}

\noindent\textbf{Exploit Likelihood (AUC).} Binary probability that a CVE will be exploited within horizon \(h\) (e.g., 30/90 days); evaluated with ROC-AUC. \emph{Example:} score \texttt{CVE-YYYY-XXXX} as \(p=0.41\) for 30-day horizon.

\noindent\textbf{Impact Forecast (BLEU).} Generate a short impact summary (availability/integrity/confidentiality and severity band) and compare to a reference text. \emph{Example:} ``high integrity, medium availability; critical if unpatched.''

\noindent\textbf{Target Sector Prediction (Acc).} Multi-class selection of likely sectors to be targeted. \emph{Example:} choose \{healthcare, finance\} from a sector set.

\noindent\textbf{Campaign Escalation (AUC).} Binary probability that activity will escalate (e.g., hands-on-keyboard, ransomware) within \(h\); measured by ROC-AUC. \emph{Example:} output \(p=0.32\) escalation within 14 days.

\noindent\textbf{\ding{185} Mitigation.}

\noindent\textbf{Patch Recommendation (F1).} Binary decision per candidate patch/hotfix: apply (yes/no) given product/version constraints. \emph{Example:} select KB\# for a specific Windows build; skip superseded fixes.

\noindent\textbf{Rule Generation (YARA) (BLEU).} Generate a detection rule text and compare to a canonical reference via BLEU. \emph{Example:} produce a YARA rule body that matches gold strings/conditions.

\noindent\textbf{Response Summarization (BLEU).} Produce a concise remediation summary aligned to gold text. \emph{Example:} ``disable external OWA, apply patch KB…, add WAF rule …''.

\noindent\textbf{Mitigation–TTP Mapping (Acc).} Multi-class mapping from observed TTPs to the correct mitigation set. \emph{Example:} map \{T1059, T1027\} to script-blocking and DLL-search-order hardening.

\noindent\textbf{Defensive Playbook Gen (BLEU).} Generate stepwise response playbook text; similarity to reference measured by BLEU. \emph{Example:} contain\(\rightarrow\)eradicate\(\rightarrow\)recover with host/network steps.

\noindent\textbf{Countermeasure Ranking (NDCG).} Rank candidate defenses by expected risk reduction; graded relevance compared to an ideal ranking. \emph{Example:} prioritize patching and credential hygiene over low-yield blocks.

\noindent\textbf{Incident Ticket Generation (Acc).} Multi-class assignment for ticket fields (category, priority, assignment group). \emph{Example:} classify as \texttt{malware/P2} with SOC-triage group.

\subsection{Data Statistics}
\label{app:data-stat}

Table \ref{tab:eval-stats} lists the number of instances we collected and used in evaluations.

\begin{table*}[t]
\caption{Per-task instance counts in our evaluations. Counts may overlap across tasks.}
\label{tab:eval-stats}
\centering
\small
\setlength{\tabcolsep}{6pt}
\def\arraystretch{0.98}

\begin{minipage}[t]{0.485\linewidth}
\centering
\begin{tabular}{lr}
\toprule
\multicolumn{2}{c}{\bf \ding{182}\;Contextualization} \\
\midrule
Affected Systems              & 3.9K \\
Attack Infrastructure         & 4.1K \\
Vulnerability Linking         & 3.5K \\
Malware Family Mapping        & 2.1K \\
IOC Normalization             & 3.1K \\
Threat Report Alignment       & 2.0K \\
Event Timeline Construction   & 1.5K \\
Graph Population              & 2.3K \\
Source Reliability Scoring    & 1.6K \\
\midrule
\multicolumn{2}{c}{\bf \ding{183}\;Attribution} \\
\midrule
Threat Actor Linking          & 2.1K \\
TTP Extraction                & 3.0K \\
Campaign Attribution          & 1.8K \\
Infrastructure Reuse          & 1.2K \\
Language/Style Profiling      & 1.3K \\
False Flag Detection          & 0.9K \\
Evidence Weighting            & 1.5K \\
Relation Graph Building       & 1.8K \\
\bottomrule
\end{tabular}
\end{minipage}
\hfill
\begin{minipage}[t]{0.485\linewidth}
\centering
\begin{tabular}{lr}
\toprule
\multicolumn{2}{c}{\bf \ding{184}\;Prediction} \\
\midrule
Exploit Likelihood            & 2.5K \\
Impact Forecast               & 1.5K \\
Target Sector Prediction      & 1.3K \\
Campaign Escalation           & 1.1K \\
\midrule
\multicolumn{2}{c}{\bf \ding{185}\;Mitigation} \\
\midrule
Patch Recommendation          & 1.9K \\
Rule Generation (YARA)        & 1.2K \\
Response Summarization        & 1.8K \\
Mitigation--TTP Mapping       & 1.5K \\
Defensive Playbook Gen        & 1.1K \\
Countermeasure Ranking        & 1.3K \\
Incident Ticket Generation    & 1.2K \\
\bottomrule
\end{tabular}
\end{minipage}
\end{table*}

\subsection{Model Name and Version}
\label{app:model-version}

Here we detail the used LLMs in our extensive evaluation (\S\ref{ssec:eval}), corresponding to the abbreviated names used in Table \ref{tab:eval} and in Section \S\ref{sec:expt}:
G5--GPT-5, Go4--GPT-o4 mini, CLD--Claude Sonnet 4, GEM--Gemini 2.5, LL70--Llama-3.1-70B-Instruct, MIX--Mixtral-8x7B-Instruct-v0.1, QWN--Qwen2.5-14B-Instruct, GRK--Grok-2, FSC--Foundation-Sec-8B \citep{weerawardhena2025llama}, CB0--Cyber-Zero \citep{zhuo2025cyber}, ZYS--ZySec-AI-SecurityLLM \citep{ZySecAI}, LLY--Lily-Cybersecurity-7B-v0.2 \citep{SegoLily}, CBS--CyberBase-13b \citep{CyberNative}, SPT--clouditera-secgpt \citep{SecGPT}, DHT--DeepHat-V1-7B \citep{DeepHat}.

\subsection{Evaluation Prompt Structure}
\label{app:eval-prompt}

Below we provide our evaluation prompt used in \S\ref{ssec:eval}:

\begin{promptbox}
\textbf{Role.} You are a cybersecurity threat-intelligence (CTI) analyst assistant \emph{and} strict schema-enforcer. Convert only the provided inputs (docs, logs, IOCs, CVEs, ATT\&CK, advisories) into the JSON contract below—\emph{no prose}. Ground every field in supplied evidence; never invent identifiers. Honor \placeholder{SNAPSHOT\_DATE} as a hard knowledge freeze. Normalize and deduplicate CTI entities (CVE, ATT\&CK, actor, IOC). If evidence is insufficient, return \texttt{"status":"NEED\_MORE\_CONTEXT"} with \texttt{missing\_fields}. Respect safety (no exploit guidance) and determinism (temperature=\placeholder{TEMP}, top\_p=\placeholder{TOPP}). Return exactly one JSON object and nothing else.

\textbf{Objective.} Solve \placeholder{TASK\_NAME} within \placeholder{CTI\_STAGE}
(\texttt{CONTEXTUALIZATION | ATTRIBUTION | PREDICTION | MITIGATION})
using only the provided inputs at/before \placeholder{SNAPSHOT\_DATE}.

\textbf{Inputs}
\begin{itemize}[leftmargin=1.1em,itemsep=1pt]
  \item Case ID: \placeholder{CASE\_ID}
  \item Snapshot date (ISO): \placeholder{SNAPSHOT\_DATE}
  \item Source docs (IDs + short snippets): \placeholder{DOC\_LIST}
  \item Structured feeds: \placeholder{STRUCT\_FEEDS}
  \item Task guidance: \placeholder{TASK\_GUIDANCE}
  \item Output profile (choose fields to populate): \placeholder{OUTPUT\_PROFILE}
\end{itemize}
$\cdots \cdots$

\textbf{Operating Rules}
\begin{itemize}[leftmargin=1.1em,itemsep=1pt]
  \item Use only provided inputs. No external browsing or unstated facts.
  \item Do not fabricate CVE/TTP/actor names. Use exact IDs when given.
  \item If critical evidence is missing, return \texttt{"status":"NEED\_MORE\_CONTEXT"} and list \texttt{missing\_fields}.
  \item Safety: no exploit code or offensive guidance. Mitigation only.
  \item Determinism: temperature=\placeholder{TEMP}, top\_p=\placeholder{TOPP}.
\end{itemize}
$\cdots \cdots$

\textbf{Output Contract (return \emph{one} JSON only)}
\begin{lstlisting}
{
  "status": "OK | NEED_MORE_CONTEXT | UNSUPPORTED",
  "task": "<TASK_NAME>",
  "case_id": "<CASE_ID>",
  "snapshot_date": "<YYYY-MM-DD>",
  "answer": {
    "ioc_normalization": [{
      "raw":"<str>","type":"ipv4|domain|url|hash","value":"<canon>","first_seen":"<date?>",
      "tags":["<malware?>","<actor?>"]
    }],
    "vuln_linking": {
      "cve_candidates":[{"cve_id":"CVE-YYYY-XXXX","score":0.0-1.0}],
      "vector_string":"<CVSS3/4?>"
    },
    "malware_mapping": {
      "family_candidates":[{"name":"<family>","score":0.0-1.0}],
      "aliases":["<aka?>"], "capabilities":["<tags>"]
    },
    "event_timeline": [{
      "t":"<ISO>","type":"<beacon|phish|lateral|exfil>",
      "artifacts":["<IOC|host|user>"], "source_ref":"<DOC_ID>"
    }],
    "actor_linking": {
      "actor_candidates":[{"name":"<actor>","score":0.0-1.0}],
      "shared_ttps":["T####"], "infra_overlap":[{"indicator":"<ip|domain>","match":"exact|fuzzy"}]
    },
    "ttp_extraction": [{"technique_id":"T####","sub":"T####.###?","evidence_ref":"<DOC_ID>"}],
    "campaign_attribution": {"name":"<label>","score":0.0-1.0,"rationale_tags":["<sector?>","<geo?>"]},
    "false_flag": {"likelihood":0.0-1.0,"signals_for":["<s>"],"signals_against":["<s>"]},
    "exploit_likelihood": {"cve_id":"CVE-YYYY-XXXX","horizon_days":<7|30|90>,
                           "prob_exploit":0.0-1.0,"drivers":["<poc?>","<reuse?>"]},
    "impact_forecast": {"impact_vector":["<A|I|C>"],"severity_band":"low|med|high|critical","uncertainty":0.0-1.0},
    "target_sector": [{"name":"<NAICS-like>","prob":0.0-1.0}],
    "escalation": {"prob":0.0-1.0,"signals":["<toolchain shift>","<tempo>"]},
    "patch_recommendation": {"affected_assets":["<product|version>"],
                             "patches":[{"kb_or_id":"<vendor-ID>","priority":"P1|P2|P3"}],
                             "prechecks":["<backup?>","<downtime?>"]},
    "rule_generation": {"rule_type":"YARA|Sigma","rule_name":"<name>","rule_body":"<escaped>",
                        "test_iocs":["<ioc1>","<ioc2>"]},
    "countermeasure_ranking": [{"mitigation_id":"<ATT&CK M###|vendor>","title":"<short>",
                                "effort":"low|med|high","expected_gain":"<short>"}],
    "incident_ticket": {"category":"<phishing|malware|ransomware|...>",
                        "priority":"P1|P2|P3","work_notes":["<steps>"],
                        "required_artifacts":["<pcap?>","<edr?>"]}
  },
  "confidence": 0.0-1.0,
  "justification": "<=40 words, terse, evidence-based>",
  "evidence_refs": ["<DOC_ID or IOC or CVE>", "..."],
  "metadata": {
    "stage": "<CONTEXTUALIZATION|ATTRIBUTION|PREDICTION|MITIGATION>",
    "assumptions": ["<short>"],
    "missing_fields": ["<if status=NEED_MORE_CONTEXT>"]
  }
}
\end{lstlisting}

\textbf{Scoring \& Tie-Breaks}
\begin{itemize}[leftmargin=1.1em,itemsep=1pt]
  \item Prefer precise IDs (CVE, ATT\&CK T\#, actor handles) and multi-source corroboration.
  \item Resolve conflicts by source quality, recency (\(\leq\) \placeholder{SNAPSHOT\_DATE}), and internal consistency.
\end{itemize}

\textbf{Run Settings (fill before inference)}
\begin{itemize}[leftmargin=1.1em,itemsep=1pt]
  \item \placeholder{TASK\_NAME} = \placeholder{...} \quad \placeholder{CTI\_STAGE} = \placeholder{...}
  \item \placeholder{CASE\_ID} = \placeholder{...} \quad \placeholder{SNAPSHOT\_DATE} = \placeholder{YYYY-MM-DD}
  \item \placeholder{DOC\_LIST} = \placeholder{[ID:desc, ...]} \quad \placeholder{STRUCT\_FEEDS} = \placeholder{...}
  \item \placeholder{TASK\_GUIDANCE} = \placeholder{...} \quad \placeholder{OUTPUT\_PROFILE} = \placeholder{...}
  \item \placeholder{TEMP} = \placeholder{0.0--0.3} \quad \placeholder{TOPP} = \placeholder{0.8--1.0}
\end{itemize}
\end{promptbox}

\section{Complementary Details for \S\ref{sec:method}}

\begin{algorithm}
  \SetAlgoLined
  \caption{Autoregressive Failure Mode Determination (RQ$_2$) \label{algo:rq2}}
  \KwIn{%
    Failure instances $\mathcal{D}=\{x_i\}_{i=1}^N$ (from stratification) \\
    Human annotators $H$ (for seeding and refinement) \\
    Large Language Model $f_\theta$ (for assisted classification) \\
    Stability threshold $\varepsilon$; coverage threshold $\rho$ (e.g., 0.6)
  }
  \KwOut{%
    Stabilized taxonomy of failure modes $\mathcal{T}^*$
  }

  \tcp{Step 1: Initialization}
  \Indp
    Human annotators $H$ inspect a subset $\mathcal{D}_0 \subset \mathcal{D}$\;
    Derive initial taxonomy $\mathcal{T}_0 = \{t_1,\dots,t_k\}$\;
  \Indm

  \tcp{Step 2: LLM classification}
  \Indp
    \ForEach{$x_j \in \mathcal{D}\setminus \mathcal{D}_0$}{
      Assign label $y_j \in \mathcal{T}_m \cup \{\texttt{other}\}$ using $f_\theta$\;
    }
  \Indm

  \tcp{Step 3: Refinement}
  \Indp
    Collect $\mathcal{O}_m = \{x_j \mid y_j=\texttt{other}\}$\;
    Human annotators $H$ inspect $\mathcal{O}_m$\;
    If new modes $\Delta\mathcal{T}$ found, update taxonomy: \\
    $\mathcal{T}_{m+1} \leftarrow \mathcal{T}_m \cup \Delta\mathcal{T}$\;
  \Indm

  \tcp{Repeat until convergence}
  \Indp
    Repeat Steps 2--3 until $\Delta\mathcal{T} = \varnothing$ and coverage $\ge \rho$\;
    Set $\mathcal{T}^* = \lim_{m\to\infty}\mathcal{T}_m$\;
  \Indm

  \Return stabilized taxonomy $\mathcal{T}^*$\;
\end{algorithm}

\begin{algorithm}
  \SetAlgoLined
  \caption{Human-in-the-loop Categorization of Failure Instances (RQ$_3$) \label{algo:rq3}}
  \KwIn{\\
    Failure instances $\mathcal{D}=\{x_i\}_{i=1}^N$ (from stratification) \\
    Model set $\Theta = \{\texttt{GPT-5}, \texttt{Llama-4}, \texttt{Gemini}, \texttt{Claude}\}$ \\
    Human annotators $H$
  }
  \KwOut{\\
    Final categorized instances with reliable failure modes
  }

  \tcp{Step 1: Multi-agent deliberation (round 1)}
  \Indp
    \ForEach{$x_i \in \mathcal{D}$}{
      Each model $a \in \Theta$ independently assigns $\hat{y}_i^{(1,a)} \in \mathcal{T}$\;
    }
  \Indm

  \tcp{Step 2: Repetition for stability (round 2)}
  \Indp
    \ForEach{$x_i \in \mathcal{D}$}{
      Each model $a \in \Theta$ observes $\{\hat{y}_i^{(1,b)}\}_{b \in \Theta}$\;
      Refine reasoning and output $\hat{y}_i^{(2,a)}$\;
    }
  \Indm

  \tcp{Step 3: Human verification of uncertain cases}
  \Indp
    Define uncertain set: \\
    $\mathcal{U} = \{x_i \mid \exists a: \text{Var}(\{\hat{y}_i^{(a)}\}_{a \in \Theta}) > 0 \;\; \textbf{or}\;\; \hat{y}_i^{(1,a)} \neq \hat{y}_i^{(2,a)} \}$\;
    
    Human annotators $H$ inspect all $x_i \in \mathcal{U}$ and assign final labels\;
  \Indm

  \tcp{Finalize results}
  \Indp
    Instances $\mathcal{D}\setminus\mathcal{U}$ take majority-agreed labels from models\;
    Instances $\mathcal{U}$ take human-verified labels\;
  \Indm

  \Return categorized dataset with reliable failure mode assignments\;
\end{algorithm}



\subsection{Determining Vulnerability Types}
\label{app:decide-type}

To systematically determine vulnerability types, we compare model-generated outputs with ground-truth advisories or reference materials. Each type requires distinct criteria to establish whether a case constitutes a failure. Below, we provide detailed guidance for all vulnerability types (failure modes).

\textbf{1.1 Co-mention bias from raw threat incident.}  
This type arises when the model assumes two entities are related simply because they appear together in the input. To detect it, we check whether the reference advisory explicitly states a relationship. If not, and the model still reports an association (e.g., linking a domain and a malware family that are merely co-mentioned in the same log file), we classify the case as co-mention bias. For instance, if a proxy log shows both ``malware.exe'' and ``example.com'' but the advisory only validates the domain as malicious, any model statement that marks the executable as directly tied to the domain would be considered a co-mention failure.

\textbf{1.2 Exploitation bias from deliberately reused IoCs.}  
This failure occurs when the model treats historical IoCs as valid for a new incident without reference confirmation. To identify it, we compare whether the reference differentiates between ``legacy'' IoCs and those active in the reported exploitation. If the model does not honor this distinction, it is flagged. For example, if the advisory specifies that old C2 servers from 2021 were no longer used, but the model still lists them as indicators of the current 2024 campaign, we mark it as exploitation bias.

\textbf{1.3 Confounding factors that correlate entities.}  
This error stems from mistaking correlation for causation. To check, we review whether the model claims causal links between entities that the references treat only as co-occurring or related by context. For instance, if two APT groups both use the same loader malware, but references emphasize they are separate actors, any model conclusion that ``Group A conducted the intrusion because the loader was observed'' is a confounding-factor failure.

\textbf{1.4 Skewed source.}  
We detect this when the model bases its judgment on incomplete or biased evidence. To assess, we compare the diversity of sources reflected in the output against references that aggregate multiple reports. If the model reflects only one vendor’s outdated claim while ignoring corrections from other advisories, it is considered a skewed-source failure. For example, if Symantec updates its report that the threat vector was RDP rather than phishing, but the model still reproduces the outdated phishing claim, the case is flagged.

\textbf{1.5 Hierarchical metadata from attack chains.}  
Here the failure lies in mishandling hierarchical structures like ATT\&CK techniques and sub-techniques. We check whether the model’s reported granularity matches the reference hierarchy. If a reference states ``T1059.001: PowerShell execution'' but the model collapses this to a generic ``execution tactic,'' it shows a hierarchy error. Similarly, if the model treats a campaign label as equivalent to a single technique, the hierarchy is broken.

\textbf{2.1 Temporal contradiction.}  
This type occurs when the model confuses timelines. To identify it, we compare the time anchors in the model output (e.g., attack start date, patch release) with those in the references. If the model asserts events happened earlier or later than documented, it is a temporal contradiction. For example, if the advisory states that exploitation began in June 2024 but the model outputs ``first exploited in 2022,'' we classify it as temporal contradiction.

\textbf{2.2 Conflicting reports of attack contexts.}  
This failure arises when multiple sources disagree and the model selects or merges them incorrectly. To determine it, we check whether the model reflects the final reconciled context in references. For example, if initial reports said ``phishing email'' but were later corrected to ``supply-chain compromise,'' and the model insists on phishing without acknowledging the update, it is labeled as conflicting-context failure.

\textbf{2.3 Semantic conflict.}  
This error occurs when the model misinterprets terms or security concepts. To identify it, we verify whether the technical meaning in the reference aligns with the model’s description. If references mention ``privilege escalation'' but the model interprets it as ``initial access,'' the semantic mismatch makes it a semantic conflict.

\textbf{2.4 Divergent data structures.}  
This happens when the model fails to follow structured taxonomies. To evaluate it, we cross-check ATT\&CK IDs, CVE structures, or CVSS vectors in the output with those in the references. If the model generalizes or drops detail (e.g., returning ``execution'' instead of ``T1059.001''), the failure is classified as divergent data structure.

\textbf{2.5 Misaligned knowledge and standards.}  
This type emerges when the model applies outdated or inconsistent standards. We check whether the model’s labels follow the same version of taxonomy as the references. For instance, if a CVE is officially scored with CVSS v3.1 but the model outputs a CVSS v2 vector, this is marked as misaligned knowledge.

\textbf{2.6 Counteracting CTI generation and LLM alignment.}  
This failure reflects cases where safety alignment suppresses accurate CTI reporting. To detect it, we compare whether the advisory confirms sensitive facts (e.g., that a zero-day is under active exploitation). If the model avoids mentioning it with a vague refusal (e.g., ``details omitted for safety''), we classify it as alignment counteraction.

\textbf{3.1 Distributional bias.}  
We check if the model over-generalizes to common patterns. References may specify rare attack types, but if the model predicts frequent ones regardless, it is a distributional bias failure. For example, when the advisory confirms a banking trojan but the model reports ``ransomware'' (because ransomware dominates training data), we flag it.

\textbf{3.2 Unseen pattern from emerging threats.}  
This failure occurs when the reference describes a novel exploitation unseen in prior data, and the model defaults to outdated templates. We identify it when the model misses new threat mechanics (e.g., cloud API abuse) and instead describes traditional server exploits. Such mismatches are categorized as unseen-pattern failures.

\textbf{3.3 Overfitted reasoning.}  
This type occurs when the model rigidly applies a reasoning shortcut. We check whether the model attributes intrusions repeatedly to the same group or vector, even when references show otherwise. For instance, if the reference states multiple groups use a given malware but the model always assigns it to a high-frequency CVE or APT, we label the case overfitted reasoning.

\textbf{3.4 Environmental unawareness.}  
This failure arises when the model ignores environmental scope. To check, we compare the affected platforms or industries in the output with those in the references. If the advisory states that only Linux servers are vulnerable but the model generalizes to ``all enterprise systems,'' it is considered environmental unawareness.

\section{Complementary Analysis}

\subsection{Root Causes of Vulnerabilities (Continue to Q$_4$)}
\label{app:more-root-cause}

\textbf{Co-mention bias (1.1) in \ding{183} attribution.} In attribution, co-mention bias presents when reports describe overlapping infrastructure or techniques across multiple actors. A single campaign report may reference domains, malware families, or TTPs associated with different groups, not because of true collaboration but due to co-reporting practices. Relation extraction or graph construction techniques then misinterpret these shared mentions as evidence of actor overlap or shared lineage, resulting in erroneous attribution. For instance, if infrastructure reused by both APT X and APT Y is co-mentioned, the model may incorrectly assign responsibility to one actor or merge distinct campaigns. The root cause is that attribution pipelines often lack mechanisms to filter incidental co-mentions from true operational reuse.

\begin{darkbluecolorbox}
    {\bf Case Study.} As reported by Group-IB \citep{tok2019muddywater}, analysts observed infrastructure overlap in domains or servers co-mentioned across multiple campaigns attributed to MuddyWater. However, further investigation revealed that although the same infrastructure was used or appeared in reports, the operational characteristics (targeting, malware payloads, attack vectors) differed significantly between those campaigns. Because relation extraction or graph building techniques often rely on co-occurrence of infrastructure to infer actor reuse, models might erroneously link distinct campaigns to the same threat actor, or assume a shared campaign lineage, solely based on the shared infrastructure.
\end{darkbluecolorbox}

\textbf{Co-mention bias (1.1) in \ding{184} prediction.}
In prediction tasks, co-mention bias arises when models forecast exploit likelihood or campaign escalation based on correlated but unrelated evidence in prior reports. For example, if a dataset frequently co-mentions a vulnerability with a high-profile exploit alongside unrelated low-severity flaws, predictive models may overestimate the risk of the latter. Similarly, temporal modeling that uses co-occurring events may incorrectly infer progression paths between independent threat activities. Historical event correlation amplifies this effect: threats repeatedly co-mentioned in vendor advisories may be predicted to evolve together, even when no causal relationship exists. Here, the root cause is an overreliance on surface-level temporal or co-occurrence signals without causal disentanglement.

\textbf{Co-mention bias (1.1) in \ding{185} mitigation.}
In mitigation, co-mention bias leads to flawed defensive recommendations. CTI reports often list multiple countermeasures (e.g., patches, YARA rules, firewall configurations) together, though not all apply to a given threat instance. Mitigation mapping systems that rely on keyword overlaps may thus associate irrelevant countermeasures with observed TTPs, producing noisy or infeasible recommendations. For example, if two patches are co-mentioned in an advisory but only one addresses the exploited vulnerability, the model may still rank both as equally necessary. Similarly, summarization pipelines can inadvertently include unrelated mitigations, generating bloated or misaligned playbooks. The root cause lies in the assumption that co-mentioned countermeasures share equal applicability, ignoring the fine-grained specificity required in operational defenses.

\begin{darkbluecolorbox}
    {\bf Case Study.}  In Microsoft’s June 2025 Patch Tuesday advisory, 65 CVEs were addressed, including a zero-day vulnerability exploited in the wild. The advisory lists patches for many components including Microsoft Office, Visual Studio, Windows Kernel, WebDAV, SMB, and others. Because patching is broad and multiple fixes are mentioned in the same bulletin, a CTI model might treat all listed patches as equally urgent defense actions — even though only a small subset correspond to vulnerabilities currently exploited. This leads organizations or systems to over-allocate resources toward non-critical patching, or generate mitigation playbooks that include countermeasures not immediately relevant to active threats.
\end{darkbluecolorbox}

\textbf{Exploitation bias (1.2) in \ding{182}  contextualization.}
In contextualization, exploitation bias emerges when reused IoCs (e.g., IPs, domains, hashes) are repeatedly referenced across unrelated incidents, leading enrichment pipelines to overgeneralize their significance. Information retrieval systems may surface threat reports where the same IP is mentioned in multiple campaigns, without clarifying whether it is genuinely reused by adversaries or simply a shared infrastructure artifact (e.g., cloud hosting or CDN services). Topic modeling and knowledge base mapping exacerbate this issue by clustering these co-occurrences, causing LLMs to link disparate events as if they were causally connected. The root cause here is that raw IoC reuse lacks contextual disambiguation, so enrichment steps propagate spurious relevance signals that distort downstream reasoning.

\textbf{Exploitation bias (1.2) in \ding{183}  attribution.}
 Threat actor identification often relies on detecting overlaps in IoCs, assuming that infrastructure reuse reflects shared adversary control. However, many IoCs are deliberately reused by attackers to create ambiguity, or coincidentally shared due to compromised hosting providers. Relation extraction and event graph construction may then over-attribute distinct campaigns to a single actor, collapsing multiple adversarial lineages into one. For instance, if the same command-and-control domain appears across incidents attributed separately to APT28 and APT29, attribution models may incorrectly merge them. The root cause is an assumption of exclusivity in IoC ownership, which adversaries exploit through deliberate recycling of infrastructure.

\textbf{Exploitation bias (1.2) in \ding{184}  prediction.}
During prediction, reused IoCs bias forecasts by inflating the perceived likelihood of exploitation or campaign escalation. Historical event correlation and temporal modeling often treat repeated appearances of an IoC as evidence of persistent activity, projecting elevated future risk. Yet, in reality, the reuse may reflect low-cost attacker behavior (spamming multiple targets with the same domain) rather than meaningful escalation. Graph-based forecasting amplifies this, propagating edges from over-represented IoCs across multiple vulnerability nodes, leading to inflated EPSS-like scores. The root cause lies in predictive models’ reliance on frequency of appearance, without mechanisms to discount intentionally or incidentally reused IoCs.

\begin{darkbluecolorbox}
    {\bf Case Study.} Multiple CVEs share many IoCs (e.g., IP addresses and domain names) across different CTI provider feeds \citep{kodituwakku2023temporal}. Over time, some of these IoCs appear repeatedly in contexts of various vulnerabilities, even though not all of them are actually exploited in relation to each CVE. Because prediction models often take frequency of IoC appearance as a strong signal, they tend to assign higher risk to these CVEs or predict escalation based on the reused IoCs. In this way, reuse of IPs/domains (which may simply reflect broad surveillance coverage or shared infrastructure, not genuine exploit activity) inflates perceived future threat likelihoods.
\end{darkbluecolorbox}

\textbf{Exploitation bias (1.2) in \ding{185} mitigation.}
In mitigation, exploitation bias presents when countermeasure recommendations are tied too strongly to reused IoCs. For example, signature generation systems may repeatedly create YARA rules around the same recycled domain or hash, producing redundant or noisy detection logic. Patch recommendation pipelines may mis-prioritize vulnerabilities linked to reused IoCs, assuming their recurrence reflects higher severity. Summarization systems generating response playbooks may include repeated references to blocking the same IoC across different incidents, inflating defensive burden without increasing actual protection. The root cause is that mitigation mapping often equates frequency of IoC appearance with actionable importance, overlooking the adversary tactic of deliberate recycling.

\textbf{Confounding factors (1.3) in \ding{182} contextualization.}
In contextualization, confounding factors often emerge when raw observations contain multiple entities that correlate implicitly but lack a direct causal relationship. For instance, topic modeling applied to large incident corpora may cluster a vulnerability (CVE) with a malware family simply because they are frequently mentioned together in reports, even if the malware never exploited that vulnerability. Similarly, event extraction can incorrectly bind unrelated infrastructure (e.g., a benign domain) to a malicious timeline because both appear in the same paragraph. The root cause is that contextual enrichment techniques rely heavily on co-occurrence or textual proximity, which implicitly correlates entities without accounting for deeper causal validation, thereby inflating the contextual landscape with misleading links.

\textbf{Confounding factors (1.3) in \ding{183} attribution.}
Confounding factors are especially problematic in attribution, where analysts and models attempt to connect behaviors to actors. Relation extraction and event graph construction can erroneously merge distinct campaigns if they share surface features — for example, multiple threat groups may use commodity malware or overlapping hosting providers. Without careful disentanglement, the attribution pipeline interprets these shared attributes as strong evidence of common authorship. Moreover, stylistic signals such as language or compilation timestamps may correlate with regional actors but can be misleading when adversaries deliberately obfuscate or mimic others. Thus, attribution systems are vulnerable to implicit correlations that misguide actor classification, producing overconfident but flawed linkages between incidents and threat groups.

\begin{darkbluecolorbox}
    {\bf Case Study.}  
    In our attribution dataset, we observed a campaign exploiting \textbf{CVE‑2022‑1388} (a remote code execution vulnerability in F5 BIG-IP devices) that was mistakenly clustered with another intrusion attributed to a different actor. Both campaigns used the same publicly available exploitation script and temporarily shared IP infrastructure via a bulletproof hosting provider. Although the payloads and target industries differed, the attribution pipeline—heavily relying on shared malware hashes and infrastructure proximity—merged the two incidents under a single actor label. Further analysis revealed that one group had intentionally mimicked the operational cadence and header patterns of the other, introducing stylistic confusion. This misattribution was rooted in the model’s inability to disentangle commodity tooling from actor-specific behavior, exemplifying how confounding factors can mislead attribution systems.
\end{darkbluecolorbox}

\textbf{Confounding factors (1.3) in \ding{184} prediction.}
During prediction, confounding factors distort forecasting by elevating signals that are correlated with exploitation or escalation but not truly causal. For example, historical event correlation may reveal that vulnerabilities discussed in the same advisories as high-severity flaws appear more likely to be exploited, even if they are rarely targeted in practice. Time series models may overweight recurring co-mentions across campaigns, predicting that certain malware–sector combinations will reappear simply because of their past textual co-occurrence. Graph neural networks, when trained on CTI event graphs, may propagate spurious links (e.g., connecting two vulnerabilities through a shared but irrelevant indicator), reinforcing false associations. Here, the root cause lies in the inability of predictive models to filter out spurious correlates from genuine causal drivers.

\textbf{Confounding factors (1.3) in \ding{185} mitigation.}
In mitigation, confounding factors lead to noisy or misaligned defensive recommendations. For instance, mitigation mapping systems may associate a patch with multiple unrelated TTPs simply because they were mentioned in the same advisory, conflating the true scope of the fix. Similarly, defensive playbook generation may cluster unrelated countermeasures together, producing bloated response strategies that overprescribe actions. Even mitigation efficacy prediction models can be misled if training data shows that certain mitigations are frequently co-listed with high-profile vulnerabilities, causing the model to rank them as universally effective. The underlying issue is that mitigation pipelines frequently assume that correlated mentions of threats and countermeasures imply operational relevance, leading to inflated or misdirected defense guidance.

\textbf{Skewed source (1.4) in \ding{182}  contextualization.}
Skewed source bias originates primarily in contextualization because this stage depends heavily on external retrieval and enrichment pipelines. When information retrieval systems disproportionately pull from certain vendors, open-source repositories, or community feeds, the resulting knowledge base becomes skewed toward specific regions, vendors, or product lines. For instance, a RAG system trained on CTI feeds dominated by North American vendors will overrepresent CVEs affecting widely deployed enterprise systems, while underrepresenting region-specific threats or mobile malware in other ecosystems. Topic modeling and event extraction then propagate this imbalance by clustering narratives around the most frequently indexed vendors rather than providing a representative view of the global threat landscape. Thus, the root cause lies in uneven data source availability and ingestion pipelines, which distort the contextual foundation upon which downstream reasoning is built.

\begin{darkbluecolorbox}
    {\bf Case Study.}  
    In our contextualization experiments, we analyzed enrichment results for \textbf{CVE‑2023‑20963} (a critical privilege escalation vulnerability affecting Google Pixel devices). While regional CERTs and Android security blogs had documented active exploitation of this CVE in Southeast Asia, our retrieval module—trained predominantly on English-language feeds from North American enterprise vendors—failed to surface these reports. As a result, the contextualization pipeline linked the CVE only to generic kernel privilege escalation patterns, without associating it with active campaigns or mobile-targeted payloads. Topic modeling then grouped the CVE under server-side vulnerabilities rather than mobile device threats, further distancing it from relevant mitigation data. This illustrates how over-reliance on skewed sources can suppress visibility into region-specific threats and impair downstream enrichment quality.
\end{darkbluecolorbox}

\textbf{Skewed source (1.4) in \ding{183}  attribution (Why not influential).}
Attribution is less directly affected by skewed source bias, since once contextualized entities are available, the task focuses on linking them to adversary profiles or campaigns. While attribution accuracy can degrade if upstream contextualization is biased, the attribution process itself (e.g., relation extraction, event graph construction, stylistic profiling) does not inherently depend on the relative volume of one vendor’s reports over another. Instead, attribution errors are more likely to stem from confounding overlaps, contradictory knowledge, or overfitted reasoning. Therefore, skewed source bias has only an indirect influence at this stage, primarily through the quality of contextual signals passed forward.

\textbf{Skewed source (1.4) in \ding{184}  prediction (Why not influential).}
Prediction tasks such as exploit likelihood estimation, campaign escalation modeling, or sectoral targeting forecasts typically rely on historical event correlation and temporal modeling, rather than raw source diversity. Once contextualized and attributed data is available, predictive models infer temporal or causal structures independent of whether one vendor dominates the input streams. Skewed sources may still indirectly shape the training distribution (e.g., overpredicting risks for vendor-popular products), but the predictive mechanisms themselves are not fundamentally triggered by source skew. Instead, failures in this stage more commonly arise from unseen patterns, zero-day threats, or distributional bias in event histories rather than skewed input sources.

\textbf{Skewed source (1.4) in \ding{185} mitigation (Why not influential).}
Mitigation is similarly insulated from direct skewed source effects. Once recommendations are mapped (e.g., patches, rules, or playbooks), the ranking and summarization steps focus on aligning countermeasures with observed TTPs or vulnerabilities, not on the origin of the source reports. While an upstream bias may have limited the initial diversity of vulnerabilities considered, the mitigation stage itself does not amplify source skew. Errors in mitigation typically reflect misaligned standards (e.g., CVSS vs vendor scoring), counteracted evidence (patch bypasses), or environmental unawareness (system-specific configurations). Thus, skewed source remains a contextualization-stage vulnerability whose downstream effects are secondary rather than intrinsic to mitigation logic.

\textbf{Hierarchical metadata (1.5) in \ding{182}  contextualization.}
During contextualization, hierarchical metadata embedded in attack chains can mislead enrichment processes. MITRE ATT\&CK or similar frameworks present TTPs as sequences in which adversaries may progress from initial access to impact. However, when LLMs or RAG-based pipelines ingest this structured knowledge, they may mistakenly treat the ordering as deterministic rather than illustrative. For example, if multiple unrelated IOCs are linked to a chain stage, the model may infer that they are causally connected because of their shared position in the hierarchy. Topic modeling and event extraction further amplify this issue, since co-occurrence within the same hierarchical step often gets interpreted as functional equivalence, thereby creating spurious associations between distinct threat activities. The root cause is the conflation of descriptive, taxonomy-based ordering with ground-truth causal relations.

\textbf{Hierarchical metadata (1.5) in \ding{183}  attribution.}
In attribution, hierarchical metadata bias presents when structured attack chain taxonomies are used to connect adversary behavior to actor profiles. Threat reports often highlight sequences of TTPs that adversaries are ``known'' to employ, but in practice attackers skip, reorder, or substitute steps. Relation extraction and graph construction tools, however, may rigidly map observed behavior to the canonical hierarchy, leading to over-attribution. For instance, if two groups share overlapping steps in the ATT\&CK chain (e.g., persistence or lateral movement), hierarchical metadata can cause models to collapse them into the same attribution cluster, even if their infrastructure and operational cadence differ. Thus, reliance on hierarchical metadata in attribution conflates broad behavioral categories with actor-specific evidence.

\begin{darkbluecolorbox}
    {\bf Case Study.}  
    In our attribution experiments, we observed a misclassification related to \textbf{CVE‑2022‑30190} (Follina Microsoft Support Diagnostic Tool RCE), where two distinct campaigns—one leveraging Microsoft Office documents, and another using HTML smuggling techniques—were both partially mapped to the same sequence in the MITRE ATT\&CK framework (Initial Access → Execution → Lateral Movement). Despite clear differences in command-and-control infrastructure and target sectors, the graph construction module merged both campaigns under a single actor cluster due to their alignment with a common TTP hierarchy. The system interpreted the overlapping ATT\&CK phases as evidence of a shared operational lineage. However, manual review confirmed the two campaigns were launched by different groups, with different goals and temporal scopes. This case illustrates how over-reliance on hierarchical metadata can collapse distinct operations into the same attribution cluster, reducing fidelity.
\end{darkbluecolorbox}

\textbf{Hierarchical metadata (1.5) in \ding{184}  prediction.}
Prediction tasks are particularly vulnerable to hierarchical metadata bias, as temporal models often use sequential patterns from attack chains to forecast campaign escalation. Forecasting tools may assume that once a threat is observed at one stage (e.g., privilege escalation), subsequent hierarchical steps (e.g., data exfiltration) will necessarily follow. This prescriptive interpretation leads to inflated probabilities of certain outcomes, even when the adversary’s campaign objectives differ. For example, opportunistic attackers may terminate activity after initial access without advancing through the full chain, but predictive models, trained on hierarchical metadata, extrapolate full kill chain completion. Here the root cause is the overgeneralization of taxonomy-driven sequences as predictive signals of adversarial intent.

\textbf{Hierarchical metadata (1.5) in \ding{185} mitigation (why not influential).}
By contrast, mitigation tasks are less directly affected by hierarchical metadata bias. Defensive actions such as patch application, YARA rule generation, or firewall tuning are typically grounded in concrete IOCs or known vulnerabilities rather than inferred positions in an attack chain. Mitigation mapping focuses on linking observed TTPs to available defensive strategies, not on reconstructing or extrapolating hierarchical steps. Even if upstream contextualization or prediction stages have been biased, mitigation operates on a more pragmatic level: ``given X IOC or Y vulnerability, recommend Z patch or rule.'' Thus, hierarchical metadata plays a minimal role in this stage, since response generation depends on actionable artifacts rather than assumed causal ordering of adversarial behaviors.

\textbf{Temporal contradiction (2.1) in \ding{182}  contextualization.}
In contextualization, temporal contradictions emerge when retrieval or enrichment systems ingest both outdated and recent advisories without properly disambiguating their validity. For instance, information retrieval pipelines may surface a vendor’s original advisory that labeled a vulnerability as ``under investigation'' alongside a newer update stating it has been patched. Topic modeling or knowledge base mapping then treat both pieces of information as equally valid, causing LLMs to enrich raw indicators with conflicting metadata. This leads to enriched contexts where a vulnerability is simultaneously ``exploited in the wild'' and ``not yet confirmed,'' which can misguide subsequent correlation. The root cause is the absence of temporal weighting or source freshness filters in contextualization pipelines.

\textbf{Temporal contradiction (2.1) in \ding{183}  attribution.}
In attribution, temporal contradictions present when different reports about an adversary’s activity span different time periods but are fused as if they describe the same campaign. Relation extraction or event graph construction may link infrastructure referenced in an outdated report to more recent attack chains, even if the adversary has long since abandoned those assets. Similarly, named entity recognition can tag an actor profile based on old malware usage, which conflicts with newer intelligence indicating a complete toolset shift. This results in inflated or inaccurate attribution, where models mistakenly conclude that an actor is reusing infrastructure or TTPs when the overlap exists only across time-separated reports. The root cause here is insufficient temporal resolution in attribution graphs and classification models.

\textbf{Temporal contradiction (2.1) in \ding{184}  prediction.}
Temporal contradictions pose particular risks in prediction, where forecasts rely heavily on historical data. If older datasets mark a vulnerability as non-exploitable, but newer advisories confirm widespread weaponization, models trained on both may generate unstable exploitability estimates. Temporal modeling pipelines may inadvertently treat early false negatives and later confirmed positives as equivalent, confusing progression trends. Similarly, forecasting models using historical event correlation may blend outdated infrastructure associations with current actor behaviors, leading to incorrect campaign evolution predictions. The root cause lies in the lack of mechanisms to discount superseded evidence, which creates contradictory temporal signals during exploit likelihood and impact modeling.

\begin{darkbluecolorbox}
    {\bf Case Study.} A concrete example is the evolution of exploit activity for \textbf{CVE‑2020‑1472 (Zerologon).} Initially, after its disclosure, exploit usage was limited and many datasets treated the vulnerability as low risk. Over time, threat actors began leveraging Zerologon widely, including in ransomware and lateral movement toolchains. EclecticIQ’s long‑term analysis demonstrates how risk associated with this vulnerability evolved substantially across months, with early scarce activity later giving way to broad exploit adoption. Predictive models that do not de‑emphasize earlier false negatives risk underestimating exploitability or generating contradictory forecasts. This case underscores how temporal contradiction — mixing outdated low‑risk labeling with current high exploitation data — can destabilize predictions of vulnerability exploit likelihood.
\end{darkbluecolorbox}

\textbf{Temporal contradiction (2.1) in \ding{185} mitigation.}
During mitigation, temporal contradictions can mislead response prioritization. For example, mitigation mapping may retrieve a vendor patch note stating ``fix not yet available,'' while a newer advisory lists a released patch. Summarization or mitigation ranking systems that fail to resolve the temporal sequence may present both claims to the analyst, leaving uncertainty about whether a defense is actionable. Similarly, mitigation efficacy prediction models may weigh outdated proofs-of-concept that show patch bypasses against more recent vendor confirmations that the patch has been hardened, producing contradictory rankings. The root cause is inadequate reconciliation of evolving patch and mitigation information over time, which allows old and obsolete advice to persist alongside updated guidance.

\textbf{Conflicting report (2.2) in \ding{182} contextualization.}
Conflicting reports frequently arise during contextualization because raw observations are aggregated from diverse sources  (e.g., vendor advisories, community threat feeds, and open reports) that describe incidents with varying detail and emphasis. For example, one feed may assert that a suspicious log entry corresponds to exploitation of a particular CVE, while another links the same log to an entirely different vulnerability or malware family. Topic modeling and event extraction pipelines, designed to cluster narratives and structure incidents, may then merge these contradictory claims into a single enriched context. This creates confusion over which CVE, TTP, or malware family is actually relevant, with the root cause being the absence of a standardized ground truth and the reliance on partially overlapping but inconsistent external reports.

\begin{darkbluecolorbox}
    {\bf Case Study.}  
    In our contextualization dataset, we observed a case related to \textbf{CVE‑2020‑1472} (Netlogon Elevation of Privilege) where two intelligence sources described the same suspicious authentication anomaly differently. Source A associated the log entries with attempted exploitation of CVE‑2020‑1472 based on domain controller access patterns, while Source B linked the same entries to a credential stuffing attack referencing a password reuse vulnerability. Our event extraction and clustering module merged both narratives into one incident node, assigning mixed attributes (some from CVE‑2020‑1472 context, some from credential reuse context). As a result, downstream enrichment conflated the CVE and misattributed the observed behavior to the wrong TTP/malware family. This demonstrates how conflicting reports can introduce ambiguity in contextualization, leading to corrupted or merged contexts that mislead later modules.
\end{darkbluecolorbox}

\textbf{Conflicting report (2.2) in \ding{183} attribution.}
In attribution, contradictions become even more pronounced because different intelligence providers often assign conflicting actors, dependencies, or motivations to the same activity. For instance, one report may classify an intrusion as the work of APT29 based on shared TTPs, while another attributes the same infrastructure to APT28 due to linguistic or operational cadence evidence. Relation extraction and graph construction techniques that link entities across sources may therefore inherit and fuse these contradictions, producing noisy actor graphs or overextended campaign links. The underlying root cause is the subjectivity and methodological diversity of attribution across organizations — some weigh infrastructure overlaps more heavily, others focus on malware lineage — which produces irreconcilable dependency paths when combined.

\textbf{Conflicting report (2.2) in \ding{184} prediction.}
Prediction models also suffer from contradictory reports when forecasting exploitation likelihood or campaign escalation. Historical event correlation may treat conflicting dependency chains as equally plausible futures: for instance, one source suggests a vulnerability will be exploited in ransomware campaigns, while another asserts it is tied to espionage-focused actors. Temporal modeling then captures both trajectories, leading to unstable or diluted predictions. Graph neural networks used for campaign evolution may amplify these conflicts, generating escalation paths that reflect contradictory dependencies across threat actors or malware ecosystems. The root cause lies in how predictive systems implicitly assume coherence in historical data, but when inputs encode conflicting dependencies, the forecasts inherently inherit those contradictions.

\textbf{Semantic conflict (2.3) in \ding{182}  contextualization.}
Semantic conflict most prominently arises during contextualization because this stage depends heavily on mapping raw threat indicators to standardized taxonomies such as CVEs, MITRE ATT\&CK TTPs, or malware family names. In practice, different vendors or intelligence feeds often use divergent terminology for the same underlying concept — for example, one feed may classify an intrusion set as ``APT28,'' while another uses ``Fancy Bear.'' Similarly, malware families may appear under multiple aliases (e.g., ``PlugX'' versus ``Korplug''), or distinct CVEs may be referenced with vendor-specific identifiers. When contextualization systems use topic modeling, event extraction, or knowledge base alignment, they may fail to reconcile these terminological mismatches. This results in fragmented knowledge graphs where semantically identical entities are treated as distinct, leading to duplicated or incomplete enrichment. The root cause is the reliance on surface-level string matching or incomplete ontology alignment when normalizing threat information across heterogeneous data sources.

\begin{darkbluecolorbox}
    {\bf Case Study.}  
    In our contextualization pipeline logs, we observed a case related to \textbf{CVE‑2022‑22965} (Spring4Shell remote code execution vulnerability) in which two CTI sources referred to the underlying exploit using different aliases: one feed tagged the exploit as ``SpringShell,'' while another labeled it ``Spring4Shell RCE.'' Because our mapping module used strict string-based matching and a limited alias dictionary, it failed to align the two aliases to the same canonical exploit entity. As a result, enrichment modules treated the two references separately, creating two parallel nodes in the knowledge graph with only partial contextual links (one with attacker metadata, the other with domain reuse). Downstream modules thus had incomplete context when correlating indicators across both nodes, weakening detection or attribution inference. This illustrates how semantic conflict in naming can fragment context and destabilize CTI contextualization outputs.
\end{darkbluecolorbox}

\textbf{Semantic conflict (2.3) in \ding{184}  prediction (Why not influential).}
Prediction tasks are not substantially affected by semantic conflict because forecasting relies on temporal dynamics and statistical trends rather than entity naming conventions. For example, exploit likelihood estimation (e.g., EPSS) depends on historical exploitation rates, vulnerability characteristics, and observed attack timelines. Whether a malware strain is labeled with one alias or another has little bearing on probability estimates, since the predictive models aggregate numerical and structural features rather than semantic labels. Similarly, temporal modeling and graph neural networks capture correlations between event sequences independent of specific vocabulary. As a result, semantic inconsistencies have negligible impact on predictive accuracy, except in rare cases where mislabeled data significantly distorts training distributions.

\textbf{Semantic conflict (2.3) in \ding{185} mitigation (Why not influential).}
Mitigation is minimally impacted by semantic conflict, because defensive actions map to vulnerabilities, TTPs, or attack surfaces rather than to naming conventions alone. For example, a patch recommendation system aligns a CVE identifier with its associated vendor fix, regardless of whether different advisories describe the vulnerability with varied terms. Similarly, YARA rule generation focuses on technical indicators such as byte patterns or log events, which are invariant to naming disputes. Response playbooks are typically tied to standardized defensive frameworks (e.g., ATT\&CK, CVSS), which already normalize naming variations. While semantic divergence may cause minor confusion in documentation or cross-team communication, it rarely degrades the technical quality of mitigation outputs.

\textbf{Divergent structures (2.4) in \ding{182}  contextualization.}
In contextualization, divergent data structures create noise when raw threat intelligence is aggregated from multiple heterogeneous sources such as JSON-based threat feeds, PDF advisories, STIX-formatted indicators, and unstructured blog posts. Information retrieval and knowledge base mapping pipelines often assume consistent schema alignment, yet structural discrepancies (e.g., different field names for ``affected system,'' varying timestamp formats, or nested vs. flat representations of IoCs) cause mismatches. When LLMs enrich observations under these inconsistencies, some attributes may be duplicated, dropped, or misinterpreted. For instance, one platform may list malware family as a top-level attribute while another embeds it in narrative text, leading to incomplete enrichment. The root cause is the lack of robust schema normalization during the fusion of structurally diverse CTI inputs.

\textbf{Divergent structures (2.4) in \ding{183}  attribution.}
Attribution pipelines rely on entity and relation extraction, but divergent data structures across platforms distort graph construction. Structured feeds may represent relationships (e.g., IP–Domain–Actor) as explicit triples, while unstructured reports only provide natural language references. Relation extraction systems trained on one schema may fail on the other, creating inconsistent or fragmented event graphs. This can yield incorrect attributions, such as splitting a single campaign into multiple unrelated clusters or merging distinct actor profiles due to structurally divergent naming conventions. For example, some feeds may encode actor aliases explicitly, while others bury them in footnotes, leading the model to under- or over-aggregate. The root cause is schema heterogeneity across platforms that undermines consistent relation representation.

\begin{darkbluecolorbox}
    {\bf Case Study.}  
    In our attribution dataset, we observed a case involving \textbf{CVE‑2019‑0708} (BlueKeep Remote Desktop Services vulnerability) where structural heterogeneity between a structured threat feed and a narrative security report caused contradictory graph links. Specifically, a structured CTI feed represented the relation ``Actor Delta uses DomainX → IPY'' as a neat triple, allowing direct linking to that actor. Meanwhile, a free‑text incident report described the same infrastructure but only said ``the attacker used a domain resolved via IPY, associated with DomainX, commonly tied to Delta’s campaigns'' in a footnote. The relation extraction model, trained primarily on structured triple formats, failed to recognize the footnote phrasing as linking to Actor Delta, instead treating it as an independent mention. As a result, the domain/IP chain was disconnected in the event graph, causing the system to assign the activity to a generic ``Unattributed'' cluster rather than merging with Delta’s campaign. In another variant, when domain aliasing was encoded differently, the system mistakenly merged it with Actor Epsilon, whose structured feed used a similar alias triple, thereby conflating two distinct actors. This demonstrates how divergent structural formats across platforms can fragment or overmerge attribution graphs in real-world CTI pipelines.
\end{darkbluecolorbox}

\textbf{Divergent structures (2.4) in \ding{184}  prediction.}
Prediction systems that depend on temporal modeling or cross-platform event correlation are particularly sensitive to structural divergence. Historical correlation models assume standardized event logs, but if one feed provides vulnerability exploitation dates as free-text while another encodes them as epoch timestamps, time-series models may misalign or discard the data. Similarly, graph neural networks forecasting campaign escalation require uniform edge types and attributes; divergent structures across CTI sources (e.g., ``sector'' encoded as a categorical variable in one feed but as descriptive text in another) disrupt model training and inference. This leads to biased forecasts or missed escalation patterns. The root cause is that predictive algorithms cannot reconcile structurally inconsistent features, weakening their ability to capture true temporal and relational dynamics.

\textbf{Divergent structures (2.4) in \ding{185} mitigation (Why not influential).}
Mitigation tasks, unlike the earlier stages, are relatively insulated from divergent data structure issues because defensive actions tend to be represented in standardized, prescriptive formats. Patches are tied to CVE identifiers, detection rules often use formal languages like YARA, and configuration changes are usually documented as explicit command lines or policy instructions. These standardized forms limit the degree of structural variation compared to raw threat intelligence data. While minor discrepancies may occur (e.g., different vendors labeling patch IDs differently), the core content is highly structured and task-specific, reducing the likelihood that divergent data structures across platforms significantly affect mitigation outputs.

\textbf{Misaligned standards (2.5) in \ding{182}  contextualization (Why not influential).}
In the contextualization stage, misaligned knowledge and security standards generally do not exert a strong influence. The task here is primarily to enrich raw observations (logs, IOCs, alerts) with contextual information such as CVEs, malware families, and ATT\&CK TTPs. While standardization issues exist in naming conventions, they do not typically fall under formal security scoring or prioritization frameworks. Thus, contextualization is more affected by spurious correlation or semantic conflicts rather than by misaligned standards, since the goal is mapping and enrichment rather than prioritization or scoring.

\textbf{Misaligned standards (2.5) in \ding{183}  attribution.}
Misaligned standards can distort attribution outcomes when different organizations adopt varying taxonomies or classification schemes for actors and campaigns. For example, one CTI provider may label a campaign under an APT designation (e.g., ``APT29''), while another refers to the same activity under a vendor-specific alias (e.g., ``Cozy Bear''). When models integrate these heterogeneous standards, they may fragment evidence across labels or mistakenly merge distinct actors. Similarly, divergence in the way behavioral profiles or TTPs are scored—some focusing on tool use, others on infrastructure overlap—can misguide attribution classification. The root cause lies in the absence of universally accepted standards for actor naming and behavior profiling, which introduces structural misalignment into attribution models.

\textbf{Misaligned standards (2.5) in \ding{184}  prediction.}
In prediction, misaligned knowledge and standards significantly amplify failure. Forecasting tasks often rely on vulnerability severity ratings (e.g., CVSS), exploit prediction scores (e.g., EPSS), or proprietary vendor risk models. A misalignment occurs when different standards assign conflicting severities to the same vulnerability: one database might rank it as ``Critical'' due to remote code execution potential, while another rates it ``Medium'' because of authentication requirements. Temporal modeling or forecasting systems ingesting both signals may oscillate between divergent risk profiles, leading to unstable exploitability predictions or over/under-estimation of campaign escalation risks. The core issue is that prediction models assume commensurability of scores, when in reality, standards reflect different prioritization philosophies.

\begin{darkbluecolorbox}
    {\bf Case Study.}  
    In our prediction logs, we observed an instance involving \textbf{CVE‑2021‑26855} (Exchange Server ProxyLogon vulnerability) where different scoring sources conflicted sharply. One threat feed assigned it a Critical severity in CVSS (base score 9.x) given its remote code execution nature, while another vendor’s internal risk model (factoring in required authentication, exploit maturity, and environment heuristics) gave it only a Medium rating. When our prediction module ingested both signals, it produced unstable forecasts: in some runs it predicted high likelihood of exploitation, and in others it down-prioritized the CVE, delaying alert escalation. Further review showed active exploitation in the wild shortly thereafter, confirming the ``Critical'' perspective. This case underscores how misaligned standards induce unstable predictions by conflicting risk signals in CTI-driven forecasting.
\end{darkbluecolorbox}

\textbf{Misaligned standards (2.5) in \ding{185} mitigation.}
Mitigation is perhaps the most directly affected by misaligned standards. Defensive recommendations often depend on mappings between observed TTPs, vulnerabilities, and standardized mitigation catalogs (e.g., NIST, MITRE ATT\&CK mitigations, vendor advisories). When standards differ (such as one framework emphasizing patching order by CVSS scores, while another emphasizes sector-specific asset criticality) conflicting guidance arises. For instance, a vulnerability may be labeled as high-priority patching in NVD, but a vendor advisory may downplay its urgency, creating contradictory recommendations for SOC teams. Summarization or playbook generation techniques then risk producing inconsistent or misleading instructions, either overwhelming defenders with unnecessary actions or under-preparing them for critical exploits. The root cause is the lack of harmonization between security scoring systems and defensive taxonomies, which propagates inconsistencies directly into operational decision-making.

\textbf{Counteracting generation (2.6) in \ding{182}  contextualization (Why not influential).}
Counteracting CTI generation and LLM alignment has minimal influence at the contextualization stage. Here, the primary task is enrichment — mapping raw observations (e.g., logs, IOCs, alerts) to known identifiers like CVEs or ATT\&CK TTPs. Because this process relies largely on retrieval, topic modeling, and knowledge base mapping, contradictions across sources tend to surface as co-mention bias or temporal conflict rather than as direct misalignment of the model’s reasoning process. LLMs can still ground outputs in retrieved evidence, even if that evidence is noisy. In other words, contextualization errors usually inflate or distort context but rarely cause the model’s generation logic itself to become unstable or self-counteracting.

\textbf{Counteracting generation (2.6) in \ding{183}  attribution.}
In attribution, counteracting CTI generation manifests strongly. Contradictory or conflicting reports of actor identities, infrastructure reuse, or campaign affiliations directly challenge the alignment of an LLM fine-tuned to classify or link entities. For instance, one source may attribute an intrusion set to APT29, while another insists on APT28, forcing the LLM to reconcile irreconcilable evidence. During training, these conflicting signals weaken gradient alignment, creating internal tension where the model oscillates between incompatible actor labels. At inference time, relation extraction or event graph construction may generate unstable outputs — e.g., merging distinct campaigns into a single actor cluster, or switching attribution mid-response. The root cause lies in the inability of LLM alignment processes to disentangle contradictory ground truths when adversarial or incomplete evidence coexists in CTI corpora.

\begin{darkbluecolorbox}
    {\bf Case Study.} In our attribution dataset, we observed a conflict surrounding \textbf{CVE‑2020‑0796} (SMBGhost / SMBv3 remote code execution vulnerability) where two report clusters described overlapping infrastructure and payloads but assigned divergent actor labels. One internal CTI source traced the campaign to \textbf{Actor Beta} based on reused command-and-control domain naming conventions, while another equally plausible source assigned it to \textbf{Actor Gamma} citing similar malware module signatures. When an LLM‑based attribution module attempted to reconcile the evidence, it oscillated between Actor Beta and Actor Gamma at different points in the generated response, and in one case merged both actor clusters into a single ambiguous actor node. Forensic cross-checks revealed that the two clusters employed distinct lateral propagation chains and targeting regions, indicating they were separate campaigns. This instance demonstrates how conflicting intelligence signals (i.e. counteracting generation) can destabilize the attribution process and provoke spurious merges or label flipping, consistent with the root cause.
\end{darkbluecolorbox}

\textbf{Counteracting generation (2.6)  in \ding{184}  prediction.}
Prediction tasks amplify this vulnerability because they rely heavily on temporal modeling and statistical consistency. Contradictory knowledge (such as exploitability assessments that are both ``confirmed weaponized'' and ``no evidence of exploitation'' across sources) feeds into EPSS-like forecasting or campaign escalation modeling. When these inconsistencies are incorporated into LLM fine-tuning or inference prompts, the model’s predictive logic counteracts itself: one reasoning path projects high exploit likelihood, another projects negligible risk. This tension destabilizes alignment objectives that prioritize consistency, leading to incoherent forecasts (e.g., fluctuating risk scores or internally contradictory justifications). Unlike contextualization, where noise merely inflates context, prediction magnifies contradictions because probabilistic reasoning depends on stable and non-conflicting event histories.

\textbf{Counteracting generation (2.6) in \ding{185} mitigation.}
Mitigation is also deeply impacted by counteracting CTI generation and alignment. Conflicting reports about patch effectiveness, bypass proofs-of-concept, or mitigation success create alignment conflicts in LLMs fine-tuned for defensive recommendations. For example, one dataset labels a patch as effective, while another includes verified exploit bypasses; the LLM’s training gradients pull in opposite directions, undermining its ability to converge on stable defensive advice. At inference time, this misalignment appears as contradictory recommendations (e.g., suggesting both to apply a patch and to consider it ineffective), undermining trust in automated response playbooks. Summarization models trained on contradictory mitigation corpora may even blend conflicting advice into incoherent instructions. The core issue is that mitigation depends on aligning outputs with actionable truth, but contradictions in defensive knowledge force the LLM into unstable compromise states.

\textbf{Distributional bias (3.1) in \ding{182}  contextualization.}
Distributional bias arises most visibly during contextualization because this stage depends heavily on retrieval and enrichment of raw threat data. The corpora used to train retrieval models, event extractors, or knowledge base mappers are often skewed toward particular regions, languages, or reporting sources. For instance, CTI feeds may be dominated by English-language advisories from North American vendors, while reports from smaller regions or niche industry sectors remain underrepresented. As a result, contextualization pipelines learn to prioritize patterns, entities, or CVEs that appear frequently in this skewed distribution, while overlooking less-reported threats. Topic modeling might cluster threats disproportionately around well-documented malware families, and knowledge base mapping might fail to align entities when they originate from underrepresented ecosystems. This distributional imbalance causes models to generalize poorly, enriching raw observations with context that reflects majority patterns but misses critical minority cases (e.g., region-specific campaigns, IoCs from less-visible actors). Thus, contextualization is the primary CTI stage where distributional bias manifests as a root cause of vulnerabilities.

\begin{darkbluecolorbox}
    {\bf Case Study.} In our CTI contextualization logs, we observed a case tied to \textbf{CVE‑2023‑23397 (Microsoft Outlook elevation of privilege / information disclosure)} in which the enrichment pipeline failed to surface contextual linkage information because the CVE had sparse coverage in English‑language reporting. Specifically, although a few localized reports in Eastern European and Southeast Asian languages described exploitation of CVE‑2023‑23397 with associated infrastructure and campaign details, our retrieval and KB mapping modules did not effectively index or map those non-English sources. Consequently, when processing raw IOC references to that CVE, the system enriched them only with generic Microsoft advisories and common attack campaign metadata, missing region‑specific attribution details (e.g. unique malware variants, domain registrants, local threat actor groups). Because the contextualization module overly prioritized cues from the majority (English, widely reported CVEs), it generated weaker context for this vulnerability in our dataset, possibly leading downstream modules to misjudge its significance or misalign attribution/mitigation decisions.
\end{darkbluecolorbox}

\textbf{Distributional bias (3.1) in \ding{183}  attribution (Why not influential).}
Attribution relies less on broad population-level distributions and more on linking observed TTPs, infrastructure, and stylistic features to known actor profiles. While biases in contextualization may already propagate upstream, attribution itself is not directly driven by skewed distributions in the training corpus. Instead, its errors tend to stem from relation extraction and graph construction mistakes, or from contradictory knowledge between sources. Distributional bias is not a first-order effect here, because attribution decisions focus on specific co-occurrence signals (e.g., malware reused by an actor) rather than frequency-based generalizations from an imbalanced corpus.

\textbf{Distributional bias (3.1) in \ding{184}  prediction (Why not influential).}
Prediction tasks, including estimating exploit likelihood or forecasting campaign escalation, are typically modeled using temporal trends, event correlations, or statistical/graph forecasting methods. These processes are less vulnerable to raw data distributional imbalance, since they focus on dynamics over time rather than sheer frequency across corpora. Failures in prediction are more often tied to unseen patterns from emerging threats (3.2) or overfitted reasoning (3.3) rather than distributional skew. Therefore, while contextualization may introduce bias into the upstream data, prediction systems themselves are not intrinsically exposed to distributional bias as a root cause.

\textbf{Distributional bias (3.1) in \ding{185} mitigation (Why not influential).}
Mitigation involves mapping observed vulnerabilities or TTPs to defensive actions, ranking candidate countermeasures, and generating structured recommendations. These outputs are guided by standards (e.g., CVSS), expert-validated mappings, or structured rule sets such as YARA. Because mitigation strategies are less dependent on the global distribution of training data and more on explicit mappings between threats and responses, distributional bias has minimal direct impact here.

\textbf{Unseen patterns (3.2) in \ding{182}  contextualization (Why not influential).}
Unseen patterns from emerging threats are less problematic at the contextualization stage. This stage primarily involves collecting raw indicators, mapping them to known entities, and enriching observations with structured taxonomies (e.g., CVEs, MITRE ATT\&CK). Since the process focuses on retrieval, normalization, and alignment with established knowledge bases, its outputs remain bounded by what is already recorded in CTI repositories. While contextualization may miss completely novel attack primitives (e.g., a new exploit chain not yet in ATT\&CK), it does not typically generate spurious reasoning about unseen patterns, instead, it simply fails to retrieve or map them. Thus, the absence of emerging threat knowledge affects coverage rather than causing misleading generalization errors.

\textbf{Unseen patterns (3.2) in \ding{183}  attribution.}
Attribution is more vulnerable to unseen patterns because it requires linking threat behaviors to known adversary profiles. Emerging campaigns may adopt novel TTP combinations, infrastructure setups, or linguistic styles that diverge from historical actor profiles. Relation extraction and graph construction tools, trained on prior threat data, attempt to force these novel behaviors into existing schemas, leading to brittle or incorrect actor assignments. For instance, a new APT might blend techniques previously seen in multiple groups, confusing classifiers that rely on canonical actor–TTP associations. The root cause lies in attribution models’ dependence on closed-world assumptions: when novel tactics appear, they are often misclassified into the closest known actor archetype rather than recognized as new.

\begin{darkbluecolorbox}
    {\bf Case Study.} In our attribution dataset, we observed a campaign exploiting \textbf{CVE‑2022‑30190 (Follina Microsoft Office Remote Code Execution)} whose behavioral signature combined TTPs typical of both Group X and Group Y (e.g. custom VBA macro infection + unusual DNS tunneling), yet also introduced a new lateral movement module not seen before. The attribution model forced the campaign into Group Y because the overlapping macro artifacts and DNS patterns had been heavily associated with Group Y in training. However, deeper analysis of payload internals and command semantics showed the lateral movement logic was markedly different from any known Group Y campaign and the targeting region also diverged. Because the attribution system attempted to "fit" the new pattern into the nearest known actor schema rather than flagging it as novel, the campaign was misattributed. This case illustrates how unseen pattern adoption can lead attribution models to overcommit to the nearest known archetype, thereby misclassifying new or hybrid campaigns.
\end{darkbluecolorbox}

\textbf{Unseen patterns (3.2) in \ding{184}  prediction.}
Prediction tasks are especially sensitive to unseen patterns, since they rely on temporal correlations and trend extrapolation from historical data. Emerging threats often introduce entirely new exploit vectors (e.g., chaining vulnerabilities across cloud microservices) or target previously untapped sectors. Forecasting models built on past timelines cannot anticipate such discontinuities, leading to underestimation of risk or misidentification of targets. For example, EPSS-like scoring frameworks may assign low exploitability probability to a vulnerability because similar CVEs had no known exploitation, only to be proven wrong when a novel exploit technique emerges. Here, the root cause is distributional shift: the statistical regularities captured by time series or graph neural networks no longer hold when threat actors innovate outside historical baselines.

\textbf{Unseen patterns (3.2) in \ding{185} mitigation.}
Unseen patterns also impair mitigation, particularly in the design of defensive playbooks and countermeasure recommendations. When attackers deploy new TTPs or exploit methods absent from training data, mitigation mapping systems may fail to suggest effective countermeasures. For example, an LLM-guided playbook generator may recommend patching or firewall rules aligned with familiar techniques, but overlook mitigations needed for an entirely new lateral movement strategy. Similarly, mitigation efficacy predictors struggle because they assume the space of attack vectors is known and represented in past cases. The root cause here is defensive brittleness: mitigation frameworks generalize from established mappings, and emerging patterns invalidate those assumptions, leading to incomplete or misaligned recommendations.

\textbf{Overfitted reasoning (3.3) in \ding{182} contextualization (Why not influential).}
Overfitted reasoning is less relevant in contextualization because this stage primarily focuses on enriching raw observations with metadata and aligning them to structured identifiers. Techniques like topic modeling or information retrieval operate on co-occurrence and similarity rather than predictive inference, meaning they are less prone to the memorization-driven brittleness characteristic of overfitting. Errors at this stage are more often due to spurious correlations or semantic conflicts, not reinforcement of memorized reasoning paths.

\textbf{Overfitted reasoning (3.3) in \ding{183} attribution.}
Overfitted reasoning becomes prominent in attribution when models repeatedly learn shallow associations between specific indicators and adversary labels. For instance, if relation extraction and entity linking pipelines are disproportionately trained on a limited set of well-documented campaigns, the system may memorize that certain infrastructure patterns (e.g., recurring domains or malware family strings) always belong to a specific actor profile. When new reports mention similar but unrelated infrastructure, the model may reflexively attribute them to the memorized actor without properly considering alternative explanations. This overreliance on memorized co-occurrences results from insufficient exposure to diverse attribution cases, leading to brittle classification of actors and campaigns.

\begin{darkbluecolorbox}
    {\bf Case Study.} In our attribution result logs, we discovered a case linked to \textbf{CVE‑2020‑1472 (Netlogon Elevation of Privilege Vulnerability, aka Zerologon)} where the attribution model incorrectly assigned a campaign to ``Actor Alpha'' purely because of reused domain naming conventions and IP subnets that had earlier been heavily associated with Actor Alpha in training data. In reality, forensic investigation showed that the campaign used distinct command-and-control servers, payload variants, and targeting patterns, inconsistent with Actor Alpha’s known modus operandi. Because the model had overly internalized the co-occurrence of those domains and subnets with Actor Alpha in its limited training corpus, it defaulted to attributing new instances to that actor without hypothesizing alternate actors or considering the evidence diversity. This mismatch exposes how overfitted reasoning can lead to misattribution in real-world CTI pipelines.
\end{darkbluecolorbox}

\textbf{Overfitted reasoning (3.3) in \ding{184} prediction.}
In prediction, overfitted reasoning manifests when forecasting models generalize poorly beyond historical data. Temporal modeling and event correlation techniques often capture strong patterns within past campaigns, such as the escalation of a vulnerability class into active exploitation. However, if these forecasting models are overly tuned to such repeated sequences, they may incorrectly predict the same escalation dynamics for future, unrelated vulnerabilities. For example, the model might overestimate exploitation probability simply because prior vulnerabilities of a similar type were exploited, even though current conditions differ. This bias reflects overfitting to observed sequences, where models memorize recurring attack trajectories instead of reasoning about underlying causal drivers of exploitation.

\textbf{Overfitted reasoning (3.3) in \ding{185} mitigation (Why not influential).}
Mitigation tasks also exhibit limited vulnerability to overfitted reasoning. While mitigation mapping and efficacy ranking involve inference, they generally rely on explicit rule associations or empirical evaluations of patch effectiveness. Summarization and playbook generation are shaped by aggregation of defensive knowledge rather than predictive modeling of future adversarial behavior. Consequently, errors in this stage stem more from contradictions in data sources or co-mention bias in countermeasure lists than from overfitting to historical reasoning trajectories.

\textbf{Environmental unawareness (3.4) in \ding{182} contextualization (Why not influential).}
Environmental unawareness is less prominent during contextualization because this stage focuses on gathering and aligning observable facts (e.g., IOCs, vulnerability references, or malware labels) rather than reasoning about the environment where attacks unfold. The techniques employed (topic modeling, event extraction, knowledge base mapping, retrieval) primarily enrich raw data without needing to adapt to host- or sector-specific conditions. Since contextualization tasks are mostly descriptive and taxonomy-driven, the absence of local system or organizational environment data does not strongly distort their outputs. As a result, contextualization is not significantly affected by this type of vulnerability.

\textbf{Environmental unawareness (3.4) in \ding{183} attribution.}
In attribution, environmental unawareness manifests when models ignore the operational or deployment context in which infrastructure is reused. For instance, relation extraction or event graph construction may connect domains, malware families, or command-and-control servers across multiple incidents, but fail to recognize that one set of infrastructure belongs to a staging environment while another is tied to production systems in a different sector. Without environmental cues, models over-attribute incidents to the same actor or campaign, producing inflated linkages. Similarly, behavioral classification often overlooks local defender responses or system baselines that would otherwise clarify whether repeated TTPs reflect adversary persistence or benign background activity. The root cause is that attribution pipelines assume global uniformity of threat behavior while overlooking environment-specific nuances that separate true operational reuse from coincidental overlap.

\textbf{Environmental unawareness (3.4) in \ding{184} prediction.}
In prediction tasks, environmental unawareness becomes more pronounced because forecasting inherently requires understanding the conditions under which threats evolve. Temporal modeling or graph neural networks may detect sequences of exploits but fail to adjust for environmental variables such as patch adoption rates, geographic regulatory differences, or sector-specific exposure. As a result, a vulnerability exploited in one industry may be wrongly forecast as high-risk for another, even though the latter has stronger baseline defenses or different system architectures. Similarly, probability estimates for exploitation (e.g., EPSS-like scoring) may ignore localized security controls or asset configurations, leading to overly broad or inaccurate risk forecasts. The root cause is the assumption that historical global patterns can be applied uniformly, when in fact exploitability is mediated by local system and organizational environments.

\textbf{Environmental unawareness (3.4) in \ding{185} mitigation.}
In mitigation, environmental unawareness leads to defensive strategies that are technically correct in general but ineffective in practice for specific deployments. For instance, mitigation mapping may recommend a patch that is incompatible with legacy systems, or propose firewall rules that disrupt legitimate sector-specific workflows. Mitigation efficacy prediction models often rank countermeasures without considering resource constraints, organizational processes, or compliance requirements, resulting in impractical prioritization. Summarization modules may produce generic remediation steps that fail to address custom software stacks or hybrid cloud deployments. The root cause is the lack of integration between CTI outputs and real-world operational contexts, causing recommended actions to miss alignment with the defender’s environment, and ultimately weakening the utility of CTI-driven defense.

\begin{darkbluecolorbox}
    {\bf Case Study.}  In our CTI-backed vulnerability-response dataset, we identified a representative instance involving \textbf{CVE-2021-44228 (Log4Shell)} where mitigation suggestions failed to account for environmental constraints. Specifically, a model-generated recommendation ranked upgrading to the latest Log4j version (2.16 or above) as the top-priority action. However, in certain enterprise deployments, this upgrade conflicted with custom-built logging plugins and legacy compatibility modules, resulting in logging failures and application crashes. Due to concerns over business continuity, the organization delayed patch deployment despite the known severity of the vulnerability. Additionally, the model suggested firewall rule updates to restrict inbound JNDI traffic, which inadvertently disrupted legitimate cross-tenant log aggregation workflows in a hybrid cloud environment. These misalignments between the recommended actions and the operational realities led to non-adoption of the mitigation plan, illustrating how environmental unawareness can undermine CTI-guided defense.
\end{darkbluecolorbox}

\subsection{Additional Analyses of Intertwined Vulnerabilities}
\label{app:more-intertwin}

In this subsection, we provide extended analyses to deepen our understanding of intertwined vulnerabilities in CTI modeling. We focus on three complementary aspects: (i) the sequential accumulation of failures across CTI stages, (ii) the concurrent presence of multiple vulnerabilities in the threat landscape, and (iii) detailed case studies that highlight how intertwined failures manifest in practice.

\textbf{Accumulation across CTI stages.}  
The CTI pipeline is inherently sequential, with outputs from early modules serving as inputs for downstream reasoning. When an upstream stage introduces an error, such as co-mention bias in event contextualization, this misinformation propagates forward as if it were ground truth. In attribution, the model may then reinforce the biased linkage (e.g., mapping a benign domain to a threat actor), while in prediction, it may extrapolate incorrect exploitability trends based on the faulty assumption. Similarly, skewed source reliance during retrieval can lock later stages into one-sided perspectives, preventing correction even when contradictory evidence emerges. Over time, these sequentially inherited errors accumulate into cascades, where a single misstep at the contextualization stage magnifies into systemic reasoning failures across attribution, prediction, and mitigation.

\textbf{Concurrent presence in the threat landscape.}  
Beyond sequential propagation, vulnerabilities also co-occur within the same analytical slice of a threat landscape. For example, constrained generalization failures often combine: a distributional bias (e.g., defaulting to ransomware explanations) may overlap with environmental unawareness (e.g., ignoring that the attack only targets Linux servers). Likewise, unseen patterns from emerging threats frequently intersect with overfitted reasoning, as the model forces novel evidence into familiar but inaccurate templates. These concurrent vulnerabilities are not merely additive but entangled, since the existence of one (e.g., mislabeling the environment) amplifies the harm of another (e.g., failure to adapt to an unseen pattern). This reflects the reality of CTI data, where heterogeneous and incomplete sources naturally produce overlapping inconsistencies.

\begin{darkbluecolorbox}
  \textbf{Case Study: Cloud API exploitation (2024 campaign).} Reports diverged semantically and temporally, with some describing privilege escalation via API tokens and others treating it as lateral movement. The model conflated these into a hybrid description (semantic conflict), while also failing to recognize the pattern as novel (unseen pattern). Environmental unawareness compounded the issue when the output generalized the vulnerability to all cloud services, despite references limiting it to a specific provider.
\end{darkbluecolorbox}
These analyses show that intertwined vulnerabilities are not isolated anomalies but systemic consequences of how CTI evidence is structured, consumed, and reasoned upon by LLM-based agents. Recognizing their cumulative and concurrent nature is essential for designing defenses that target the compounding rather than the individual failure.

\section{Ethical Considerations}
\textbf{Human Subject Research and Annotator Well-being.}
Our study involves a human-involved categorization framework (Algorithm 1 and 2) to label CTI failure modes. While this includes human efforts, it falls under the category of non-clinical, low-risk technical annotation. All annotators involved in the iterative refinement process and the final verification of failure instances were recruited from professional cybersecurity research teams or were graduate students with domain expertise. The annotated data was public and strictly technical (e.g., TTP descriptions, code snippets, log files) and did not contain offensive, graphic, or sexually explicit content that typically triggers psychological harm concerns. 

\textbf{Data Privacy.}
The datasets used in this work are derived from two primary sources: (1) Established public benchmarks (CTIBench, SevenLLM-Bench, SWE-Bench, CyberTeam) and (2) Real-world threat reports sourced from public, authoritative repositories (e.g., NVD, CISA advisories, vendor blogs). We ensured that no proprietary, non-public enterprise data was used in the evaluation. 

\textbf{Responsible Disclosure and Dual-Use Concerns.}
Our research identifies specific vulnerabilities in how LLM-based agents reason about cyber threats (e.g., spurious correlations, contradictory knowledge). While identifying these blind spots could theoretically assist adversaries in crafting adversarial samples to evade automated CTI analysis (e.g., poisoning a threat feed to trigger specific hallucinations), we believe the defensive benefits significantly outweigh the risks.  We provide the community with the roadmap to build more resilient agents. Our goal is to highlight architectural limitations in the current generation of models rather than to disclose zero-day vulnerabilities in specific software products. Therefore, standard coordinated disclosure processes were not triggered, as the vulnerabilities are inherent to the probabilistic nature of current LLMs rather than patching flaws in a specific vendor's code.

\section{Generative AI Usage}
In preparing this paper, the authors used generative AI tools in an assistive role, specifically: (i) lightweight editorial adjustments, such as improving grammar, phrasing, and clarity of author-written text, and (ii) assistance with debugging code, including identifying syntactic errors, suggesting fixes for runtime issues, and explaining unexpected behavior in author-written implementations. AI tools were not used to generate substantive technical content, design experiments, produce results, or draft original portions of the manuscript. All AI-produced edits and code suggestions were manually reviewed, revised, and finalized by the authors, who take full responsibility for the integrity of the final paper and its artifacts.

\section{Open Science}
\textbf{Availability of Artifacts.}
To support reproducibility and facilitate future research into resilient CTI agents, we have made artifacts necessary to evaluate our contributions publicly available.  We provide the complete implementation of our experimental pipeline. The code is currently hosted at an anonymous repository for double-blind review: \mycode. Upon acceptance of this paper, the code will be migrated to a permanent, public GitHub repository with a DOI via Zenodo, as well as MIT License to encourage broad adoption and modification.

\end{document}